\documentclass{article} % For LaTeX2e
\usepackage{iclr2025_conference,times}

\usepackage{amsmath,amsfonts,bm}

\def\eqref#1{equation~\ref{#1}}
\def\1{\bm{1}}

\def\ra{{\textnormal{a}}}

\def\rx{{\textnormal{x}}}

\def\rva{{\mathbf{a}}}

\def\erva{{\textnormal{a}}}

\def\ervx{{\textnormal{x}}}

\def\rmA{{\mathbf{A}}}

\def\vmu{{\bm{\mu}}}
\def\vtheta{{\bm{\theta}}}
\def\va{{\bm{a}}}

\def\ve{{\bm{e}}}

\def\vx{{\bm{x}}}

\def\eva{{a}}

\def\mA{{\bm{A}}}

\def\mH{{\bm{H}}}
\def\mI{{\bm{I}}}
\def\mJ{{\bm{J}}}

\def\mX{{\bm{X}}}

\def\mSigma{{\bm{\Sigma}}}

\DeclareMathAlphabet{\mathsfit}{\encodingdefault}{\sfdefault}{m}{sl}
\SetMathAlphabet{\mathsfit}{bold}{\encodingdefault}{\sfdefault}{bx}{n}
\newcommand{\tens}[1]{\bm{\mathsfit{#1}}}
\def\tA{{\tens{A}}}

\def\tX{{\tens{X}}}

\def\gE{{\mathcal{E}}}

\def\gG{{\mathcal{G}}}

\def\gV{{\mathcal{V}}}

\def\sA{{\mathbb{A}}}
\def\sB{{\mathbb{B}}}

\def\sS{{\mathbb{S}}}

\def\emA{{A}}

\newcommand{\etens}[1]{\mathsfit{#1}}

\def\etA{{\etens{A}}}

\newcommand{\E}{\mathbb{E}}

\newcommand{\R}{\mathbb{R}}

\newcommand{\KL}{D_{\mathrm{KL}}}
\newcommand{\Var}{\mathrm{Var}}

\newcommand{\Cov}{\mathrm{Cov}}
\newcommand{\normltwo}{L^2}
\newcommand{\normlp}{L^p}

\newcommand{\parents}{Pa} % See usage in notation.tex. Chosen to match Daphne's book.

\usepackage{hyperref}
\usepackage{url}

\usepackage{comment}
\usepackage{xcolor}
\usepackage{graphicx}
\usepackage{duckuments}
\usepackage{caption}
\usepackage{subcaption}
\usepackage[capitalize,noabbrev]{cleveref}
\usepackage{algorithm}
\usepackage{algorithmic}
\usepackage{booktabs}
\usepackage{multirow}
\usepackage{enumitem}
\usepackage{soul}
\usepackage[normalem]{ulem}               % to striketrhourhg text
\usepackage{threeparttable}
\usepackage{bbm}
\usepackage{amsthm}
\setlist[itemize]{leftmargin=0.2in}
\newcommand\redout{\bgroup\markoverwith
{\textcolor{red}{\rule[0.5ex]{2pt}{0.8pt}}}\ULon}
\newtheorem{theorem}{Theorem}
\newtheorem{remark}{Remark}

\title{Exponential Topology-enabled Scalable Communication in Multi-agent Reinforcement Learning}

\author{
Xinran Li\textsuperscript{1,2}
\quad Xiaolu Wang\textsuperscript{3}\thanks{Corresponding author.}
\quad Chenjia Bai\textsuperscript{2}
\quad Jun Zhang\textsuperscript{1} \\
\textsuperscript{1}The Hong Kong University of Science and Technology \\
\textsuperscript{2}Institute of Artificial Intelligence (TeleAI), China Telecom \\
\textsuperscript{3}Software Engineering Institute, East China Normal University \\
\texttt{xinran.li@connect.ust.hk, xiaoluwang@sei.ecnu.edu.cn} \\
\texttt{baicj@chinatelecom.cn, eejzhang@ust.hk}
}

\iclrfinalcopy % Uncomment for camera-ready version, but NOT for submission.
\begin{document}

\maketitle
% We introduce ExpoComm, a scalable communication protocol that leverages exponential topologies for efficient information dissemination among many agents in large-scale multi-agent reinforcement learning.
\vskip -0.2in
\begin{abstract}

In cooperative multi-agent reinforcement learning (MARL), well-designed communication protocols can effectively facilitate consensus among agents, thereby enhancing task performance. Moreover, in large-scale multi-agent systems commonly found in real-world applications, effective communication plays an even more critical role due to the escalated challenge of partial observability compared to smaller-scale setups. In this work, we endeavor to develop a scalable communication protocol for MARL. Unlike previous methods that focus on selecting optimal pairwise communication links—a task that becomes increasingly complex as the number of agents grows—we adopt a global perspective on communication topology design. Specifically, we propose utilizing the exponential topology to enable rapid information dissemination among agents by leveraging its small-diameter and small-size properties. This approach leads to a scalable communication protocol, named ExpoComm. To fully unlock the potential of exponential graphs as communication topologies, we employ memory-based message processors and auxiliary tasks to ground messages, ensuring that they reflect global information and benefit decision-making. Extensive experiments on large-scale cooperative benchmarks, including MAgent and Infrastructure Management Planning, demonstrate the superior performance and robust zero-shot transferability of ExpoComm compared to existing communication strategies. The
code is publicly available at \url{https://github.com/LXXXXR/ExpoComm}.\looseness=-1

\end{abstract}

\section{Introduction}
% \textcolor{blue}{1.5 page}
% MARL -> communication in MARL -> scalability to many agents -> 

% few focus on this scenario 
% fully-connected and learn the connectivity 

% -> challenges: too expensive, not effective 

% focus too much on local and communication cost scale 

% quadratically with the number of agents 

% one-step 
% multi-step

Cooperative multi-agent reinforcement learning (MARL) has recently emerged as a promising approach for complex decision-making tasks across diverse real-world applications, such as resource allocation~\citep{marl_dis}, package delivery~\citep{marl_delivery}, autonomous driving~\citep{MARL_autonomous_driving}, robot control~\citep{marl_robot_swamy2020scaled}, and 
% infrastructure management planning
infrastructure management planning~\citep{benchmark_IMP}. Under the widely adopted centralized training and decentralized execution (CTDE)
paradigm~\citep{ctde_kraemer2016multi,ctde_lyu2021contrasting}, algorithms like MADDPG~\citep{MADDPG}, COMA~\citep{COMA}, MATD3~\citep{MATD3}, QMIX~\citep{QMIX}, and MAPPO~\citep{MAPPO} have achieved notable success.

To enhance agent collaboration in partially observable scenarios, communication mechanisms have been incorporated into multi-agent systems (MASs) to assist in decentralized decision-making~\citep{CommNet}.
% facilitate better cooperation. 
Enabling information exchange during execution helps MARL algorithms to address non-stationarity and partial observability prevalent in these environments. Building upon this foundation, researchers have devoted efforts to designing effective communication protocols, focusing on three core considerations:
1) \textit{whom} the agents should communicate with~\citep{i2c,Commformer};
2) \textit{when} communication should occur~\citep{ETC,schedNet};
and 3) \textit{how} the agents should design and utilize the communication messages effectively~\citep{TarMAC,MASIA}.
By leveraging tools such as attention and graph neural networks (GNNs), learnable and adaptive communication mechanisms have significantly advanced MARL performance. \looseness=-1

% Nevertheless, most of the previous works primarily focus on developing sophisticated communication protocols , paying less attention to the actual applicability 
% tested small-scale MASs with the number of agents less than ten \textcolor{blue}{cite papers here}, 
% leaving scalability of the communication strategies are largely overlooked. 
% Motivated by this research gap, this work focuses on the scalability of communication protocol in MARL. The challenges 
% Motivated by this research gap, this work aims to develop a scalable communication protocol in MARL that is both highly effective and low-cost.
% To address this research gap of scalability in MARL communication protocols, the challenges are two-fold: First, with many agent in the system, how . Second, how to manage the cost of the communication during execution?
% Therefore, to enable feasible applications of MARL communication protocols to real-world MASs, the research gap in scalability must be addressed. 
% To scale the current method, which involves learn the connectivity among agents, to many-agent systems, two challenges lie in the way: First, it becomes 
% In such many-agent systems, 
% Second, \underline{with a fixed sparsity of the connectivity graph typical of such methods} {\color{blue}(not very clear)}, the communication cost overall scales quadratically with \redout{respect to} the number of agents, which can be \redout{infeasibly} {\color{red}prohibitively} large in \redout{large-scale} {\color{red}many-agent} systems.
% with the increasing number in the systems 
Despite considerable success, most existing communication strategies are designed for small-scale MASs~\citep{MADDPG,smac,facmac_mamujoco} and may struggle as systems scale to dozens or even hundreds of agents, which are ubiquitous in real-world applications~\citep{population_survey,traffic_scalibility,inventory_scalibility,large_scale_networked_MARL}. 
In these \textit{many-agent} systems, existing methods that learn pairwise connectivity among agents falter for two reasons: First, these methods often require agents to receive messages only from ``useful'' peers. However, identifying these peers becomes increasingly challenging as the number of agents grows, potentially compromising the effectiveness of communication protocols~\citep{MASIA}.
Second, the overhead of these methods scales poorly. Specifically, training memory consumption quickly becomes prohibitively large, as shown in our empirical evaluation, and the communication overhead during execution scales quadratically with the number of agents, which is infeasible for many-agent systems. \looseness=-1

This motivates a fundamental rethinking of scalable MARL communication: Can we adopt a global perspective and design an overall topology that propagates information among all agents effectively and at low cost, rather than relying on finding task-specific pairwise connectivity? In this vein, we propose an exponential topology-enabled communication protocol, termed \textit{ExpoComm}, as a scalable solution for MARL communication. Unlike previous works that seek to identify useful communication links at each timestep, ExpoComm draws inspiration from graph theory and leverages the small-diameter property of exponential topologies to ensure effective communication by facilitating message flow across all agents within a limited number of timesteps. The inherent sparsity (small size) of exponential topologies allows ExpoComm's communication cost to scale (near-)linearly with the number of agents. Moreover, to fully leverage the small-size and small-diameter properties of exponential graphs for efficient information dissemination, we employ memory-based blocks for message processing and auxiliary tasks to ground messages, ensuring that they effectively reflect global information. Extensive experiments across twelve scenarios on large-scale benchmarks, including MAgent~\citep{benchmark_Magent} and Infrastructure Management Planning (IMP)~\citep{benchmark_IMP}, demonstrate the superior performance of ExpoComm compared to baseline algorithms when handling large numbers of agents up to a hundred. Additionally, owing to its global perspective without pairwise reliance, ExpoComm exhibits remarkable zero-shot transferability to larger numbers of agents during test time. \looseness=-1

\section{Related Work} \label{sec: related_work}
% \textcolor{blue}{1 page}
\paragraph{Communication in MASs}
% many communication work -> many progress for better performance -> some consider realistic concerns -> scalability -> primary issue is the communication topology -> fixed / dynamic (distance-based) -> 
% individually controlled (sender/receiver)
% GNN-learnt  

% First introduced by \citet{CommNet,DIAL_RIAL}, communication among agents in MARL has been an active research area due to its great potential to enhance cooperation and improve task performance. Owing to the high flexibility of communication protocols, finding the most effective protocols tailored to the MARL paradigm is a challenging task~\citep{MARL_comm_survey}.
% % To this end, extensive later works adopt the straightforward approach to learn the communication protocol in an end-to-end fashion, which is termed ``learning to communicate''. Notably, . M
% To this end, extensive works take the learning perspective and optimize the communication components~\citep{BiCNet}, such as message generators, message aggregators, and connectivity among agents, through end-to-end training. From the sender side, ToM2C~\citep{ToM2C} and MAIC~\citep{MAIC} enhanced the message generation process by utilizing teammate modeling. CACL~\citep{CACL} focuses on the decentralized training paradigm and proposes to learn communication encoding with contrastive learning. From the receiver side, TarMAC~\citep{TarMAC}, G2ANet~\citep{G2ANet} and MASIA~\citep{MASIA} improve message aggregation by employing attention-based aggregation strategies.

Communication among agents in MARL was first introduced by \citet{CommNet,DIAL_RIAL} and has since become an active research area due to its potential to enhance cooperation and improve task performance. The flexibility of communication protocols makes finding effective solutions for the MARL paradigm challenging~\citep{MARL_comm_survey}.
To address this difficulty, many studies have focused on optimizing communication components, such as message generators, message aggregators, and connectivity among agents, through end-to-end training~\citep{BiCNet}. From the sender side, ToM2C~\citep{ToM2C} and MAIC~\citep{MAIC} enhance message generation through teammate modeling, while CACL~\citep{CACL} uses contrastive learning techniques to learn communication encoding in a decentralized training paradigm. From the receiver side, TarMAC~\citep{TarMAC}, G2ANet~\citep{G2ANet}, and MASIA~\citep{MASIA} improve message aggregation using attention-based strategies. \looseness=-1

Recently, researchers have addressed challenges posed by real-world communication systems. Notably, NDQ~\citep{NDQ} and TMC~\citep{TMC} reduce communication costs by crafting succinct messages, while ATOC~\citep{atoc}, IC3~\citep{ic3}, I2C~\citep{i2c}, and CommFormer~\citep{Commformer} manage overhead by pruning unnecessary communication links. Additionally, \citet{noisy_channel} propose a stochastic encoding/decoding scheme to handle noisy channels, and DACOM \citep{DACOM} introduces delay-aware communication to account for the high latency of wireless channels.

Despite these advancements, scalability in communication mechanisms has been largely overlooked, often due to the quadratically increasing communication cost associated with fully-connected graphs as the number of agents grows. Although few works explicitly address the scalability issue, efforts to design communication topologies among agents offer potential solutions. These can be categorized into fully-connected, rule-based, and learned topologies. 
Early works~\citep{CommNet,DIAL_RIAL,BiCNet} typically adopt fully-connected topologies to demonstrate communication benefits, but at the cost of high bandwidth requirements. Later on, to reduce the overall communication overhead, \citet{DGN} and \citet{neighbor_graph} restrict communication to nearby neighbors based on distance, while NeurComm~\citep{NeurComm} limits communication to neighboring agents in networked MASs. In spite of achieving significant performance gains, their further applicability may be limited since they require extra information beyond local observation to determine the communication topology. In contrast, learned topology methods assume no such requirements and offer high flexibility. In particular, ATOC~\citep{atoc}, IC3~\citep{ic3}, I2C~\citep{i2c} locally deploy gates for agents to decide if they should engage in communication. However, these methods may result in uncontrollable overall communication costs due to individual control schemes. Alternatively, MAGIC~\citep{MAGIC} utilizes graph attention mechanisms to learn the communication topology, while CommFormer~\citep{Commformer} extends the idea and enables control over the overall communication sparsity. Although effective in small-scale MASs, peer-wise connectivity becomes increasingly difficult to learn in large-scale MASs, and high sparsity may impair performance, as discussed by \citet{Commformer}.
% In particular, VBC~\citep{VBC}, ETC~\citep{ETC} and MBC~\citep{MBC_sparse_comm} 

% Our proposed ExpoComm incorporates a rule-based topology, supplementing the aforementioned progress on MARL communication by explicitly addressing the scalability of communication.
Our proposed ExpoComm, which incorporates rule-based topologies for rapid information dissemination among all agents, complements existing efforts in MAS communication by explicitly addressing scalability challenges. \looseness=-1

% Subsequent works then made concrete progress 

% {\color{red}
% \textbf{Our Contributions.}
% \begin{enumerate}[label=\arabic*),leftmargin=14pt]
%     \item 
    
%     \item  
    
%     \item 
% \end{enumerate}}
\paragraph{Exponential Graphs}
% \textcolor{blue}{Should I put this paragraph above the last one?}
Exponential graphs are a class of graph topologies that exhibit strong scalability properties with respect to the number of nodes. They have been primarily used in distributed learning to periodically synchronize model updates across devices. \citet{exp_graph_decentralized_learning_first} investigate exponential graphs with gossip algorithms and achieve high consensus rates for decentralized learning. Follow-up works~\citep{SlowMo,exp_graph_decentralized_exact_avg,expG_consensus,ExpG_largebs} build upon this topology, optimizing model weight update algorithms and providing empirical evidence and theoretical guarantees for the effectiveness of exponential graphs. Beyond distributed learning, exponential graphs also have applications in chip design~\citep{wang2014rpnoc,ExpG_chip_design}. Overall, exponential graphs demonstrate efficient information dissemination across many nodes, making them a promising candidate topology for achieving scalable communication in MARL. \looseness=-1

\section{Scalable Communication with Exponential Graph in MARL} \label{sec: method}

In this section, we propose ExpoComm, which leverages exponential graphs as communication topologies among agents in MARL to enable scalable communication. We structure the following subsections to address three key questions: 1) Why and how should exponential graphs be adapted for agent communication? 2) How can the corresponding neural network architecture be designed to effectively utilize the messages transmitted through these topologies? 3) How can messages propagated among agents be grounded to ensure their usefulness?

In \cref{sec: expG_as_comm_topologies}, we outline the requirements for scalable communication: effective information dissemination among agents and low communication overhead. We translate these requirements into the challenge of identifying topologies with small diameters and sizes, key properties of exponential topologies. In \cref{sec: network_design}, we discuss how memory-based message processors can enable meaningful message encoding, leveraging the small-diameter property over multiple timesteps within exponential topologies. In \cref{sec: training_details}, we adopt a global perspective to ground messages using a global state reconstruction auxiliary task and contrastive learning, as ExpoComm aims to facilitate message flow across the entire graph rather than focusing on local features.

\subsection{Exponential Graph as the Communication Topology} \label{sec: expG_as_comm_topologies}

\subsubsection{Problem Setting} 
In this work, we consider a fully cooperative partially observable multi-agent task, which can be modeled as a decentralized partially observable Markov decision process (Dec-POMDP)~\citep{pomdp_oliehoek2016concise}. The Dec-POMDP is defined by a tuple $\mathcal{M} = \langle \mathcal{S}, A, P, R, \Omega, O, N, \gamma \rangle$ with $N$ being the number of agents and $\gamma \in (0, 1]$ being the discount factor. 
At each timestep $t$, with the global observation $s^t \in \mathcal{S}$, agent $i$ receives a local observation $ o_i^t \in \Omega$ and then communicates with other agents. Upon receiving the messages from other agents, agent $i$ then selects an action $a_i^t \in A$ based on its local policy $\pi_i$. These individual actions collectively form a joint action $\boldsymbol{a}^t \in A^N$, leading to a transition to the next global observation $s^{t+1} \sim P(s^{t+1}| s^t, \boldsymbol{a}^t)$ and inducing a global reward $r^t = R(s^t, \boldsymbol{a}^t)$. The team objective is to learn the policies that maximize the expected discounted cumulative return $G_t = \sum_t \gamma^t r^t$. 

\subsubsection{Communication Topologies} \label{sec: comm_graph_desiderata}

To design an effective and scalable communication protocol in many-agent systems, it is essential to determine whom to communicate with, i.e., to construct the communication topology so that communication is both beneficial for decision-making and cost-effective. While previous work~\citep{Commformer} assumes a static communication topology, we adopt a more flexible, time-varying directed graph $\textstyle \gG^t = \langle \gV, \gE^t \rangle$, where node $\textstyle v_i \in \gV$ denotes agent $i$ and edge $e_{i \rightarrow j}^t \in \gE^t$ indicates a communication link from agent $i$ to agent $j$ at timestep $t$. 

From a graph perspective, we consider the following desiderata for the communication topology:

\begin{enumerate}[label=$\bullet$,leftmargin=14pt]
    \item \textbf{Small graph diameter for fast information dissemination}: Formally defined as $\textstyle \text{diameter}(\gG^t) =  \max_{v_i, v_j \in \gV} d(v_i, v_j)$ with $d(v_i, v_j)$ representing the shortest path distance from $v_i$ to $v_j$, the graph diameter indicates how quickly messages travel through the graph. Since communication aids multi-agent decision-making by providing the locally observant agents with global information and alleviating the non-stationarity, a graph with a small diameter can expedite message exchange and is therefore desirable.
    % \item \textbf{Small size for low communication overhead}: Formally defined as $\textstyle \lvert \gE^t \rvert$, the size of a graph denotes the total number of edges, corresponding to communication overhead in an MAS. Given the high hardware requirement for communication modules and the potential delays induced by densely connected communication topologies, we prefer graphs with a small size in many-agent settings. 
    \item \textbf{Small size for low communication overhead}: Formally defined as $\textstyle \lvert \gE^t \rvert$, the size of a graph denotes the total number of edges, corresponding to the number of communication links in an MAS. We assume that any message transmission incurs the same overhead, therefore the total overhead scales with the number of links. Given the high hardware requirement for communication modules and the potential delays induced by densely connected communication topologies, we prefer graphs with a small size in many-agent settings. 
\end{enumerate}

\subsubsection{Exponential Graphs} \label{sec: exp_graph}

Based on the desiderata above for the communication topologies, we draw inspiration from graph literature and choose exponential graphs~\citep{exp_graph_decentralized_learning_first,exp_graph_decentralized_exact_avg} as a promising candidate for communication topology in many-agent systems. Below, we introduce two variants of exponential graphs and demonstrate their small-diameter and small-size properties through an illustrative example.

\begin{figure}[t]
  \centering
  \begin{subfigure}[t]{.21\textwidth}
    \centering
    \includegraphics[width=\textwidth]{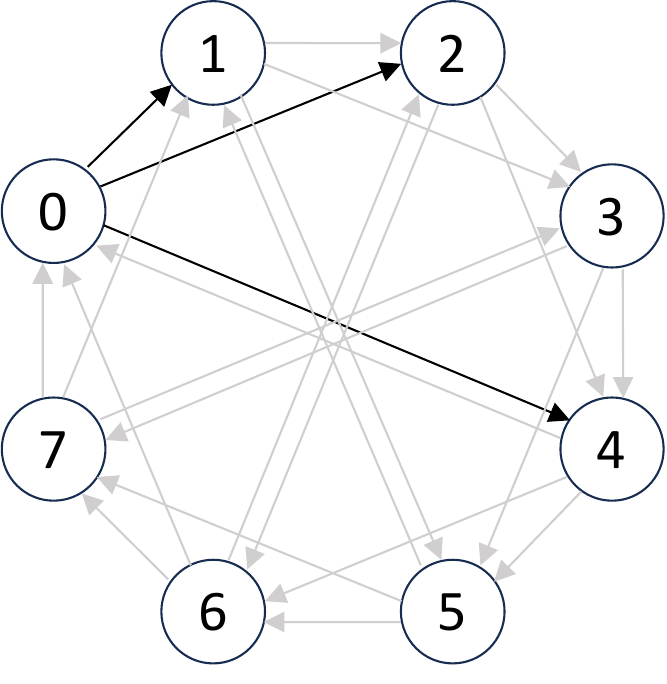}
    \caption{Static exponential graph.}
    \label{fig: demo_static_exp_graph}
   \end{subfigure}
   \hfill
  \begin{subfigure}[t]{.67\textwidth}
    \centering
    \includegraphics[width=\textwidth]{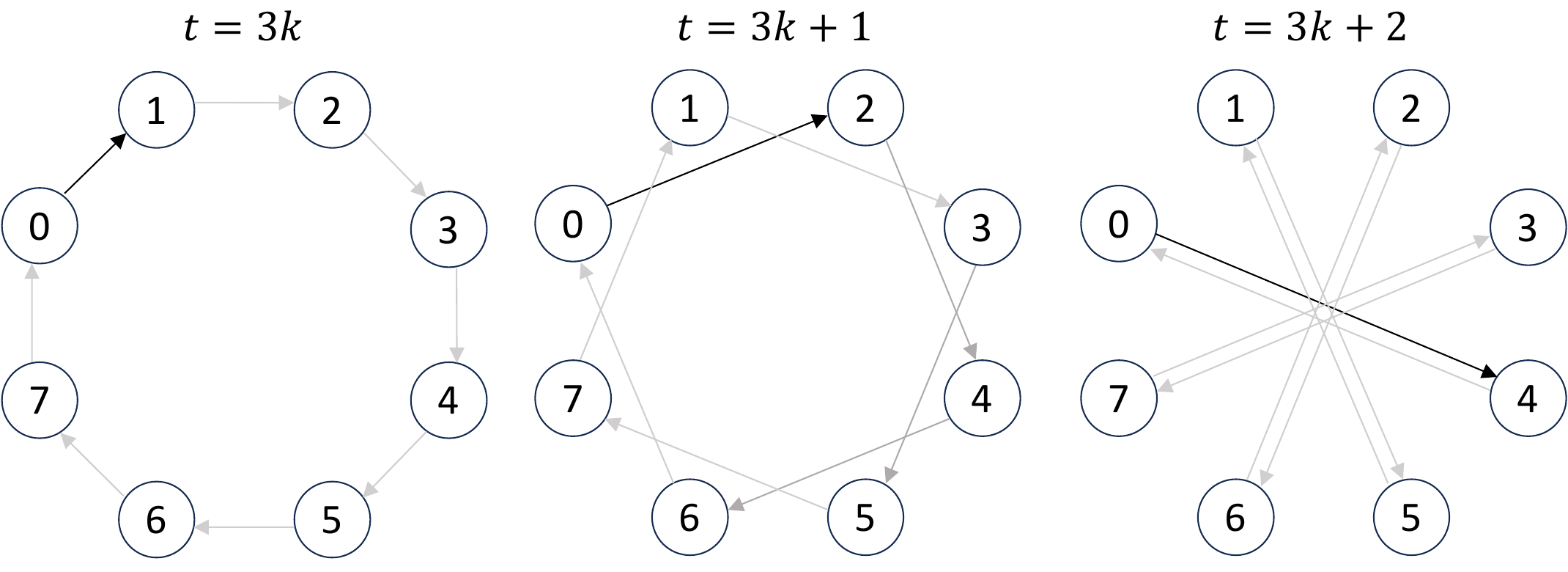}
    \caption{One-peer exponential graph. $k$ is an integer.}
    \label{fig: demo_one_peer_exp_graph}
   \end{subfigure}
  \vskip -0.1in
  \caption{Illustration of exponential graphs with $N=8$.}
  \label{fig: demo_exp_graph}
  \vskip -0.2in
\end{figure}

\paragraph{Static Exponential Graph}
Assuming a randomly sequential ordering of agents $\textstyle 0, 1, \ldots,  N-1$ and the corresponding adjacency matrix $\textstyle E \in \{0,1 \}^{N \times N}$, in the static exponential graph, each agent communicates with peers that are $\textstyle 2^0, 2^1, \ldots, 2^{\lfloor \log_2{(N-1)} \rfloor}$ hops away, which is illustrated by \cref{fig: demo_static_exp_graph}. Formally, we have
\begin{align}
    \displaystyle
    E^{t(\text{stat})}_{ij}  =
    \begin{cases}
    1 & \text{if} \log_2\left( (j-i) \bmod N \right) \text{is an integer or } i = j\\
    0 & \text{otherwise}
    \end{cases}
    .
    \label{eq: static_exp_adj}
\end{align}

\paragraph{One-peer Exponential Graph}
In the one-peer exponential graph, each agent iterates through different peers that are $\textstyle 2^0, 2^1, \ldots, 2^{\lfloor \log_2{(N-1)} \rfloor}$ hops away, which is illustrated by \cref{fig: demo_one_peer_exp_graph}. Formally, we have
\begin{align}
    \displaystyle
    E^{t(\text{one-peer})}_{ij}  =
    \begin{cases}
    1 & \text{if} \log_2\left( (j-i) \bmod N \right) = t \bmod \lfloor \log_2{(N-1)} \rfloor \text{or } i = j \\
    0 & \text{otherwise}
    \end{cases}
    .
    \label{eq: one_peer_exp_adj}
\end{align}

\begin{figure}[t]
  \centering
  \begin{subfigure}[t]{.48\textwidth}
    \centering
    \includegraphics[width=\textwidth]{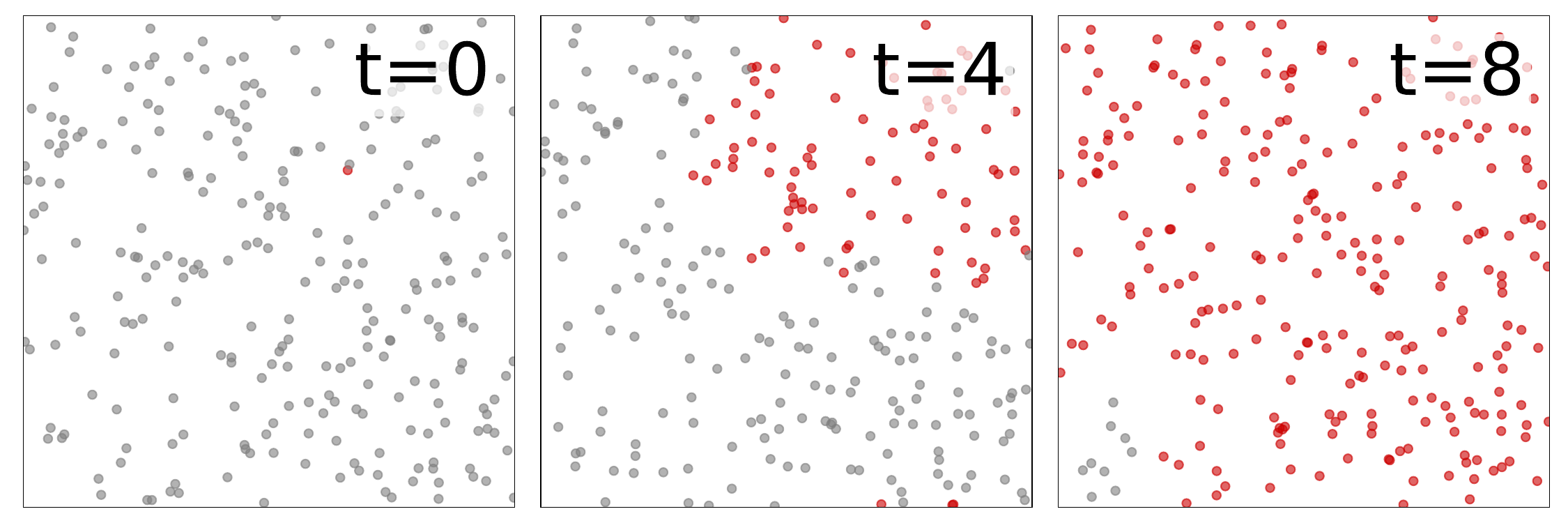}
    \caption{Distance-based graph with $\textstyle \lvert \gE^t \rvert = N \cdot \log_2N$.}
    \label{fig: demo_dense_dist}
   \end{subfigure}
   \begin{subfigure}[t]{.48\textwidth}
    \centering
    \includegraphics[width=\textwidth]{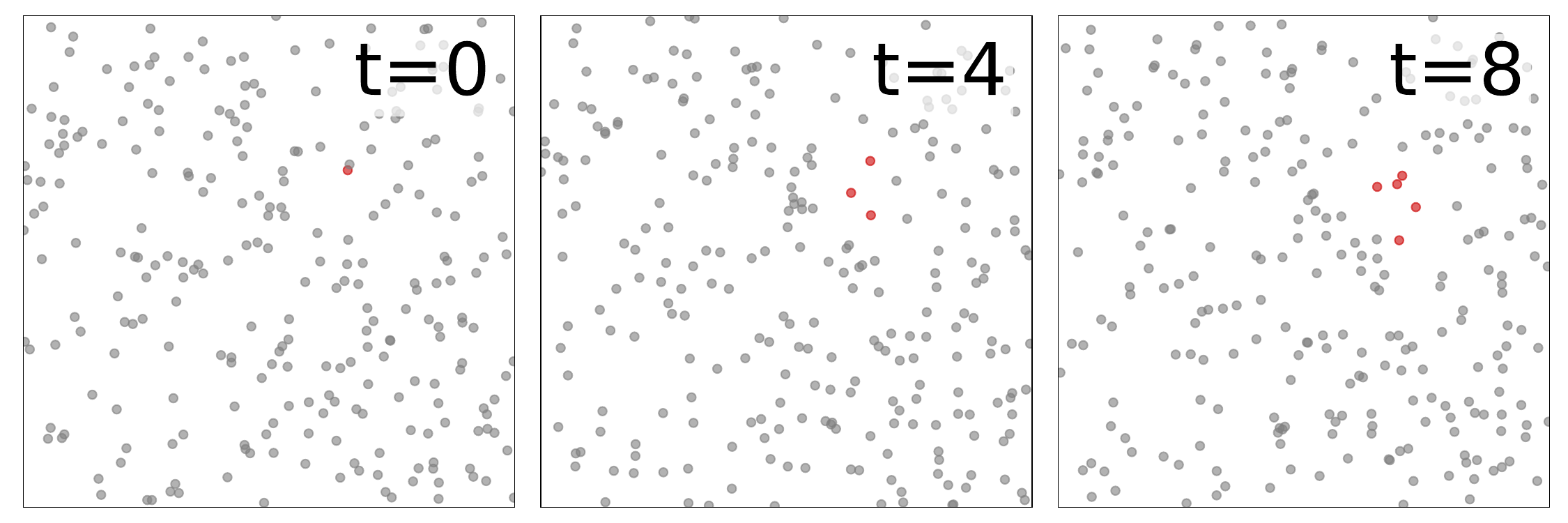}
    \caption{Distance-based graph with $\textstyle \lvert \gE^t \rvert = N$.}
    \label{fig: demo_sparse_dist}
   \end{subfigure}
   \begin{subfigure}[t]{.48\textwidth}
    \centering
    \includegraphics[width=\textwidth]{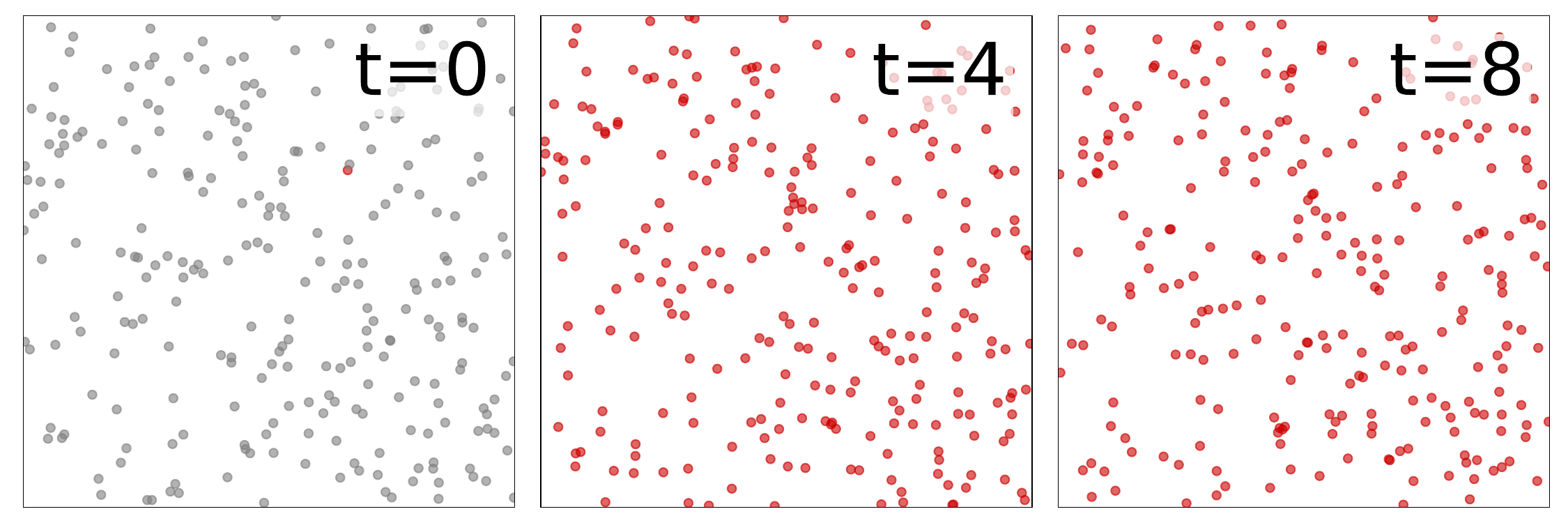}
    \caption{Erdős–Rényi graph with $\textstyle \lvert \gE^t \rvert = N \cdot \log_2N$.}
    \label{fig: demo_dense_ER}
   \end{subfigure}
   \begin{subfigure}[t]{.48\textwidth}
    \centering
    \includegraphics[width=\textwidth]{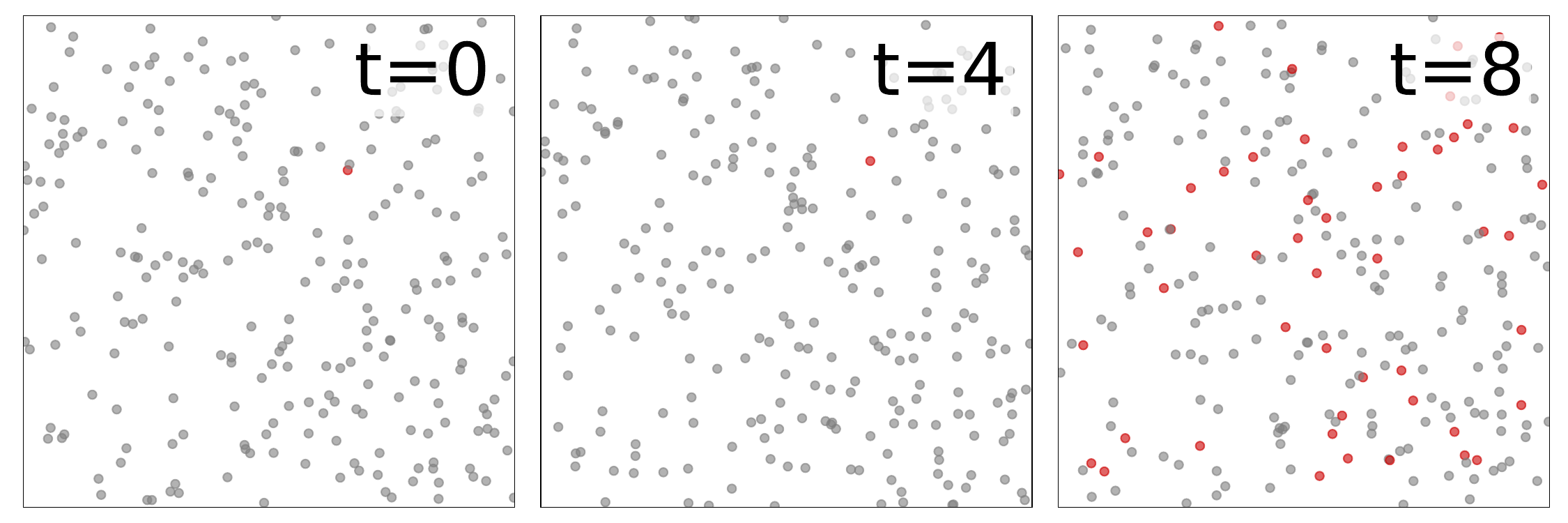}
    \caption{Erdős–Rényi graph with $\textstyle \lvert \gE^t \rvert = N$.}
    \label{fig: demo_sparse_ER}
   \end{subfigure}
   % \hfill
  \begin{subfigure}[t]{.48\textwidth}
    \centering
    \includegraphics[width=\textwidth]{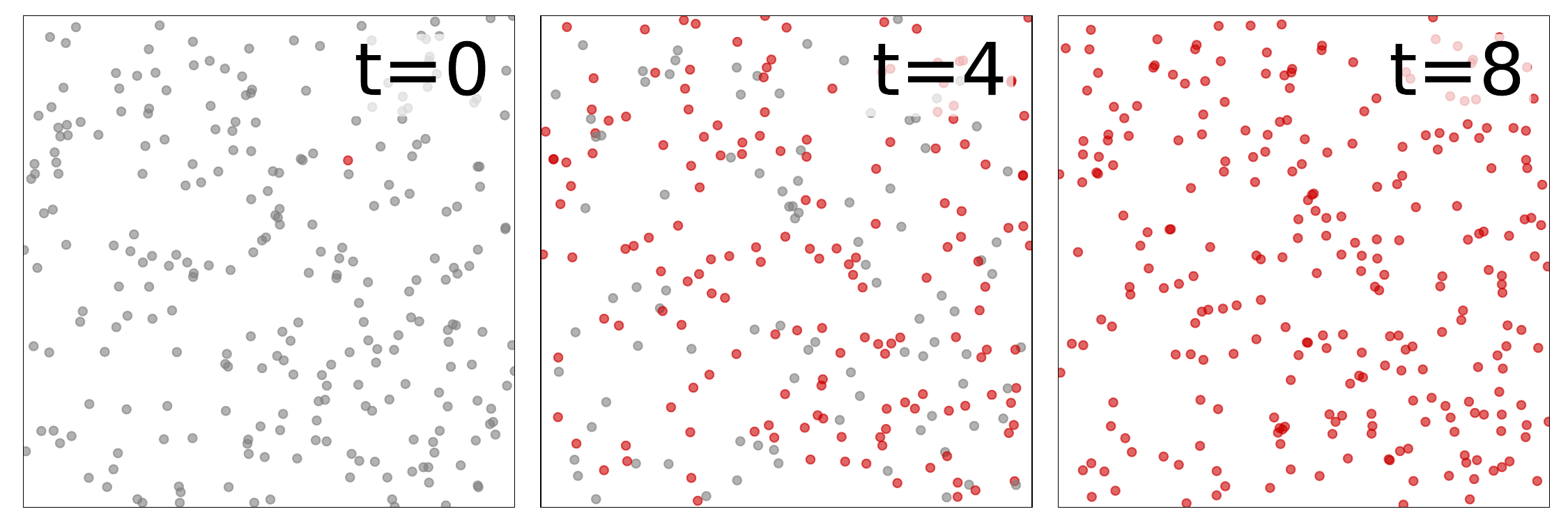}
    \caption{Static exponential graph with $\textstyle \lvert \gE^t \rvert = N \cdot \log_2N$.}
    \label{fig: demo_static_exp}
   \end{subfigure}
   \begin{subfigure}[t]{.48\textwidth}
    \centering
    \includegraphics[width=\textwidth]{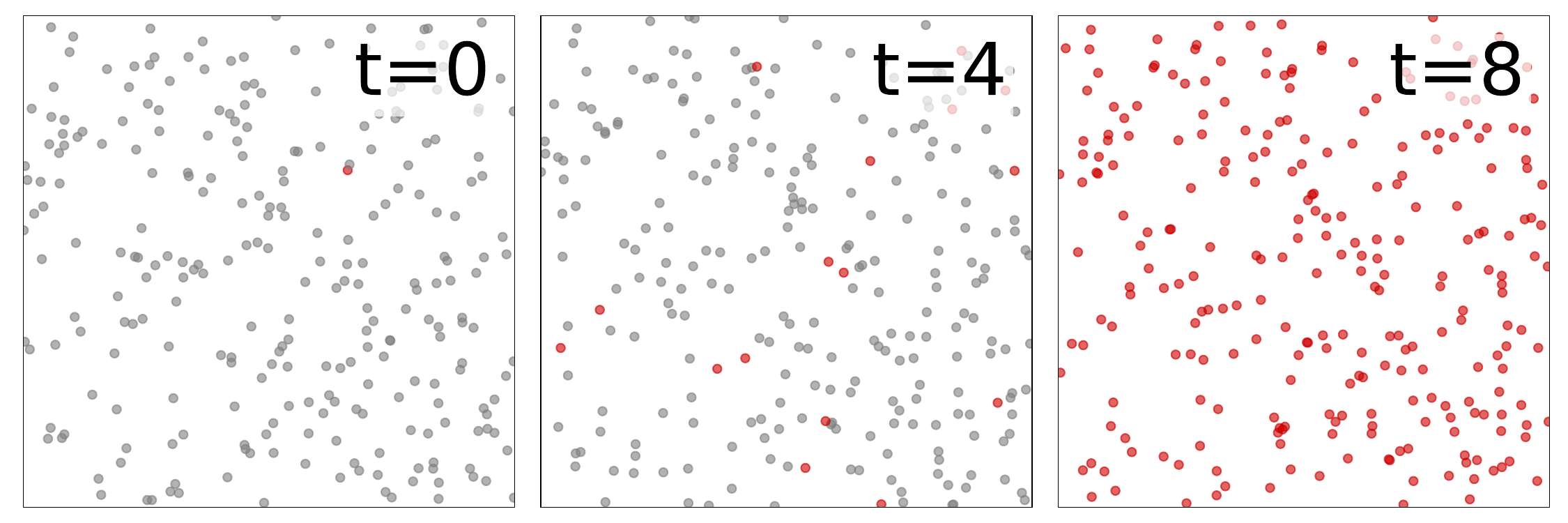}
    \caption{One-peer exponential graph with $\textstyle \lvert \gE^t \rvert = N$.}
    \label{fig: demo_one_peer_exp}
   \end{subfigure}
  \caption{A toy example to illustrate the message dissemination with different graph topologies. We demonstrate how the messages, represented by red dots, travel from a random agent to other agents over time, following different graph structures. In distance-based graphs~\citep{DGN}, agents are connected to top-$K$ nearest neighbors. In Erdős–Rényi random graphs~\citep{ER_graph}, the adjacency matrices are sampled uniformly from all the graphs satisfying the diameter and size conditions. In exponential graphs, the adjacency matrices follow \cref{eq: static_exp_adj,eq: one_peer_exp_adj}. \looseness=-1}
  \label{fig: toy_graph_diameter}
  \vskip -0.2in
\end{figure}

\paragraph{Properties}

Using the adjacency matrices defined above, we verify that the graph diameter for both static and one-peer exponential graphs is $\lceil \log_2{(N-1)} \rceil$ (see \cref{supp: theory} for details). As discussed in \cref{sec: comm_graph_desiderata}, a small diameter facilitates efficient information dissemination, especially when the number of agents $N$ is large. 

Regarding communication costs, static exponential graphs have a size of $N \cdot \lfloor \log_2{(N-1)} \rfloor$, while one-peer exponential graphs have a size of $N$. Notably, the size of one-peer exponential graphs scales linearly with the number of agents, meaning the overall communication overhead also scales linearly.

To illustrate these properties, we provide a toy example in \cref{fig: toy_graph_diameter}. We visualize the message dissemination abilities of different communication topologies under varying communication budgets. In this example, with $N=256$ agents, graph sizes (communication budgets) $\textstyle \lvert \gE^t \rvert$ are set to $N \cdot \log_2N$ and $N$, respectively.
In \cref{fig: toy_graph_diameter}, we observe that for each communication topology, reducing the graph sizes (as shown in \cref{fig: demo_sparse_dist,fig: demo_sparse_ER,fig: demo_one_peer_exp}) typically slows down dissemination speed due to increased graph diameters. This illustrates a trade-off between graph diameter and size, reflecting the trade-off between communication performance and overhead in many-agent systems. Sparser graphs with smaller sizes result in slower message dissemination but lighter communication overhead. However, exponential topologies strike a balance in this trade-off, demonstrating strong information diffusion even with a minimal communication budget of $N$.
% From \cref{fig: toy_graph_diameter}, we observe that under communication budget $N \cdot \log_2N$, both Erdős–Rényi random graphs and static exponential graphs demonstrate fast message dissemination speed. Under low communication budgets $N$, only one-peer exponential graphs show strong abilities to diffuse information among all agents. 
% Nevertheless, exponential topologies find a sweet spot in such trade-off and show strong abilities to diffuse information among all agents, even under an extremely low communication budget of $N$.

Based on these observations, we conclude that exponential topologies are well-suited for many-agent communication because: 1) In exponential topologies, any two agents can exchange messages in at most $\lceil \log_2{(N-1)} \rceil$ timesteps, ensuring timely information exchange in decentralized decision-making problems. 2) The communication overhead scales nearly linearly with the number of agents, which is crucial for many-agent systems. 3) With a rule-based topology, exponential graphs are easy to deploy and adapt to systems with varying numbers of agents, as empirically verified in \cref{sec: results}.

\subsection{Neural Network Architecture Design} \label{sec: network_design}
\begin{figure}[t]
  \centering
    \centering
    \includegraphics[width=0.8\textwidth]{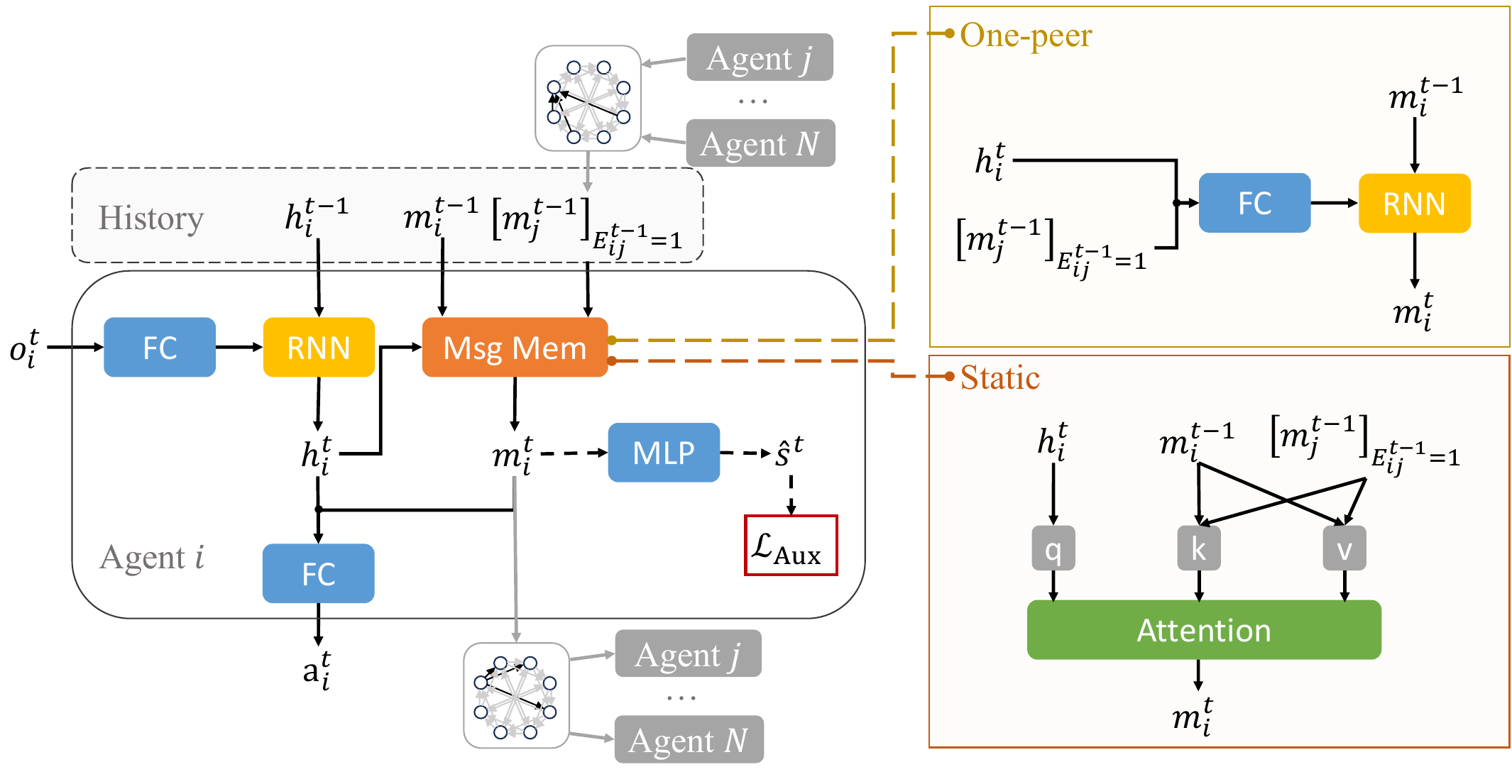}
    \caption{Neural network architecture for ExpoComm. For the static exponential topologies, attention blocks are used for message aggregation. For the one-peer exponential topologies, RNN blocks are used for message aggregation. }
    \label{fig: network_arch}
    \vskip -0.2in
\end{figure}

With exponential graphs serving as the communication topology in ExpoComm, we we elaborate on the neural network architecture to help agents utilize received messages for better decision-making. The overall architecture is illustrated in \cref{fig: network_arch}. 
ExpoComm is based on the concept of facilitating message flow across all agents within a certain timeframe, where the graph diameter indicates the length of such timeframe. To capitalize on the small graph diameter of exponential graphs, the message-processing module at each agent should ideally preserve all information received within $\textstyle \text{diameter}(\gG^t)$ timesteps. However, preserving all messages across multiple timesteps is not memory-efficient, so we employ sequential neural networks, such as attention blocks and recurrent neural networks (RNNs), for message processing.

\subsection{Training and Execution Details} \label{sec: training_details}
% \textcolor{blue}{need to revise this paragraph because the update on my method}
Following the QMIX~\citep{QMIX} algorithm, we update the network parameters $\theta$ with the objective of minimizing the temporal difference (TD) error loss:
\begin{equation}
    \displaystyle
    \mathcal{L}^{\text{TD}}(\theta) = \mathbb{E}_{(s^t, \bm{o}^t,  \bm{a}^t, r^t, s^{t+1}, \bm{o}^{t+1}) \sim \mathcal{D}} \left[\left(y^{tot} - Q_{tot}(s^t, \boldsymbol{o}^t, \boldsymbol{a}^t; \theta) \right)^2\right],
\end{equation}
where $\textstyle y^{tot} = r + \gamma \max_{\boldsymbol{a}} Q_{tot}(s^{t+1}, \boldsymbol{o}^{t+1}, \boldsymbol{a}; \theta^-)$ and $\theta^-$ represents the parameters of the target network as in DQN.

% Nevertheless, solely with the MARL training objective, 
% Moreover, since the communication 
However, communication inevitably enlarges the policy space, making it more challenging to find the optimal policy relying solely on the MARL training objective~\citep{comm_aux_loss}. To facilitate learning meaningful messages, we introduce auxiliary tasks to restore global information from local messages. From a message perspective, we aim for it to traverse among agents over multiple timesteps, accumulating new information along the way, and ultimately reflecting global information useful for decision-making.

\paragraph{Message grounding with the global state}
In scenarios where the global state is available during training, the auxiliary loss is given by the prediction error of the current global state: 
\begin{align}
    \displaystyle
    \mathcal{L}^{\text{Aux}}_{\text{pred}}(\theta, \phi) = \mathbb{E}_{(s^t, \bm{o}^t) \sim \mathcal{D}} \left[ s^t - f(m_i^t; \phi))^2 \right],
    \label{eq: aux_pred}
\end{align}
where the learnable auxiliary network for prediction $\textstyle f(\cdot; \phi)$ is used to ground the messages and can be discarded after training.

\paragraph{Message grounding without the global state} Alternatively, when the global state is unavailable during training, we use contrastive learning for meaningful message encoding, similar to \citet{CACL}. Specifically, we treat messages from different agents at the same timestep as positive pairs and messages with intervals larger than $\textstyle \text{diameter}(\gG^t)$ as negative pairs, encouraging local messages $m_i^t$ to reflect the current global latent state. The corresponding auxiliary loss is given as the InfoNCE loss~\citep{InfoNCE}: 
\begin{align}
    \displaystyle
    \mathcal{L}^{\text{Aux}}_{\text{cont}}(\theta) = - \mathbb{E}_{i, j, t, t'} \left[ \log{\frac{\text{exp} \left( g(m_i^t) \cdot g(m_j^t) / \tau \right) }{\sum_{m \in \mathcal{M}} \text{exp} \left( g(m_i^t) \cdot g(m) / \tau \right)}} \right],
    \label{eq: aux_cont}
\end{align}
where $i$ is uniformly sampled from $\{0, \ldots, N\}$, $j$ is uniformly sampled from $\{0, \ldots, N : j \neq i\}$, $\mathcal{M} = \{m_k^{t'}: k \in \{0, \ldots, N\} ,t' \notin [t-\text{diameter}(\gG^t), t+\text{diameter}(\gG^t) ] \} \cup \{m_j^t\}$ with $|\mathcal{M}| = M+1$ and $m$ is uniformly sampled from $\mathcal{M}$. $g(\cdot)$ is the normalization function, $M$ is the hyperparameter indicating the number of negative pairs and $\tau$ is the temperature hyperparameter.
The overall training loss is:
\begin{align}
    \displaystyle
    \mathcal{L}^{\text{TD}}(\theta) = \mathcal{L}^{\text{TD}}(\theta) + \alpha \cdot \mathcal{L}^{\text{Aux}}(\theta; \phi),
    \label{eq: training_loss}
\end{align}
where $\alpha$ is the hyperparameter and $\mathcal{L}^{\text{Aux}}(\cdot)$ is the auxiliary loss defined by \cref{eq: aux_pred} or \cref{eq: aux_cont}, depending on whether global information is available during training. The training and execution procedures are summarized in \cref{algo: ExpoComm_training}.

\begin{algorithm}[t]
\caption{Training and Execution Procedure of ExpoComm}
\begin{algorithmic}[1]

\STATE {\bfseries Init:} Network parameters $\theta$, $\phi$, $\mathcal{D} = \emptyset$, $\text{step} = 0$, $\theta^- =\theta$
\WHILE{$\text{step} < \text{step}_\text{max}$}
    \STATE $t=0$. Reset the environment. 

    % \STATE {\bfseries Init:} Local history  
    \FOR{$t = 1, 2, ..., \text{episode\_limit}$} 
        \STATE \textit{// Decentralized execution at agent $i$}
        \STATE Update local history $h^t_i$ based on current observation $o_i^t$ and previous history $h^{t-1}_i$
        \STATE Update agent $i$'s message $m_i^t$ based on previous local message $m^{t-1}_i$ and previously received messages $\left[ m_j^{t-1} \right]_{E_{ij}^{t-1}=1}$
        \STATE \textit{// Communication}
        \STATE Send message $m_i^t$ to peers $\{j \mid E_{ij}^t=1\}$ \hfill\COMMENT{$\triangleright$ \cref{eq: static_exp_adj,eq: one_peer_exp_adj}}
        \STATE \textit{// Action, which can happen concurrently with communication}
        \STATE Sample action $a_i^t$ based on current history $h^t_i$ and current message $m_i^t$ 
        \STATE Interact with the environment $(s^{t+1}, \boldsymbol{o}^{t+1}, r^t) = \text{env}.\text{step}(\boldsymbol{a}^t)$ 

        \STATE Save the experience $\mathcal{D} = \mathcal{D} \cup (s^t, \boldsymbol{o}^t, \boldsymbol{a}^t, r^t, s^{t+1}, \boldsymbol{o}^{t+1})$
    \ENDFOR
    \STATE At some interval, update network parameters $\theta$, $\phi$ and $\theta^-$ \hfill\COMMENT{$\triangleright$ \cref{eq: training_loss}}
\ENDWHILE
\STATE {\bfseries Output:} Policy networks parameters $\theta$
% \RETURN{} Parameters for exploitation networks $\zeta$
\end{algorithmic}
\label{algo: ExpoComm_training}
\end{algorithm}

\section{Experimental Results}
% \textcolor{blue}{3.5 pages}

In this section, we evaluate ExpoComm on two large-scale multi-agent benchmarks: MAgent~\citep{benchmark_Magent} and Infrastructure Management Planning (IMP)~\citep{benchmark_IMP}. All experiments are averaged over five random seeds and
the shaded areas represent the $95\%$ confidence interval. Details on network architecture and the training hyperparameters are available in \cref{supp: net_arch_hyperparams}.

\subsection{Experimental Setups} \label{sec: experiment_setups}
\paragraph{Environment descriptions}
In this section, We test ExpoComm and baselines across twelve scenarios in two large-scale benchmarks, with the number of agents ranging from 20 to 100. Specifically, MAgent is a particle-based gridworld environment representative of the typical MARL gaming benchmarks. To expand the variety of tasks, we also include the IMP benchmark, with tasks oriented from real-world applications. More details regarding the environment settings are provided in \cref{supp: env_details}. \looseness=-1

\paragraph{Communication Budgets}
Denoting the number of agents each agent communicates to by $K$, for each baseline in each scenario, we test two communication budgets: $K = \lceil\log_2N\rceil$ and $K = 1$, where $N$ is the number of agents in the systems. 

\paragraph{Baselines}
In the following, we compare our proposed ExpoComm with four baselines: (i) \textit{IDQN/QMIX}~\citep{QMIX}: Base algorithms without communication; (ii) \textit{DGN+TarMAC}~\citep{DGN,TarMAC}: Position-based communication topologies in which agents communicate with their nearest neighbors and use TarMAC structure to aggregate messages; (iii) \textit{ER}: ExpoComm with the exponential graph topologies replaced by random graph communication topologies following the Erdős–Rényi model; (iv) \textit{CommFormer}~\citep{Commformer}: Learned communication topologies using GNN. 
For ExpoComm, we use the static exponential graph variant for $K = \lceil\log_2N\rceil$ and the one-peer exponential graph variant for $K = 1$. For DGN+TarMAC, agents communicate to top-$K$ nearest neighbors. For ER, communication graphs are sampled uniformally from all the $K$-in-regular directed graphs. CommFormer uses constraints with varying sparsity levels for different communication budgets. Official implementations of these baselines are utilized wherever available; otherwise, we closely follow the descriptions from their respective papers, integrating them into the base algorithms. More implementation details can be found in \cref{supp: implement_details}. \looseness=-1
% For each baseline in each scenario, we test the cases with budgets for communication links of $N \cdot \log_2N$ and $N$, where $N$ is the number of agents in the systems. More details on implementation can be found in \cref{supp: implement_details}.

\subsection{Results} \label{sec: results}

% ，often out performing its counterpart method ExpoComm with static 

\begin{figure}[t]
\centering
% \subfigure[]{
% \includegraphics[width=0.7\textwidth]{Styles/figs/results/0515_legend_cropped.pdf}
%     }8
\raisebox{-\height}{\includegraphics[width=0.82\textwidth]{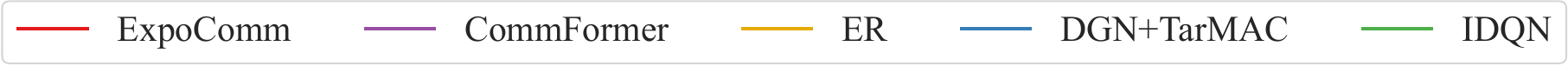}}
\par
\begin{subfigure}[t]{.32\textwidth}
    \centering
    \includegraphics[width=\textwidth]{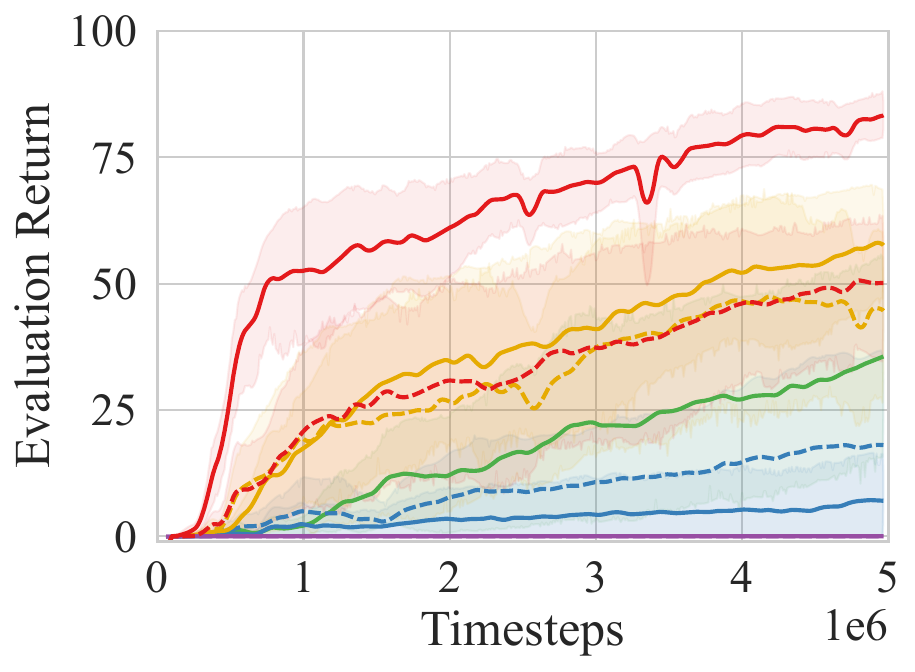}
    \caption{AdversarialPursuit w/ 25 agents}
\end{subfigure}
\begin{subfigure}[t]{.32\textwidth}
    \centering
    \includegraphics[width=\textwidth]{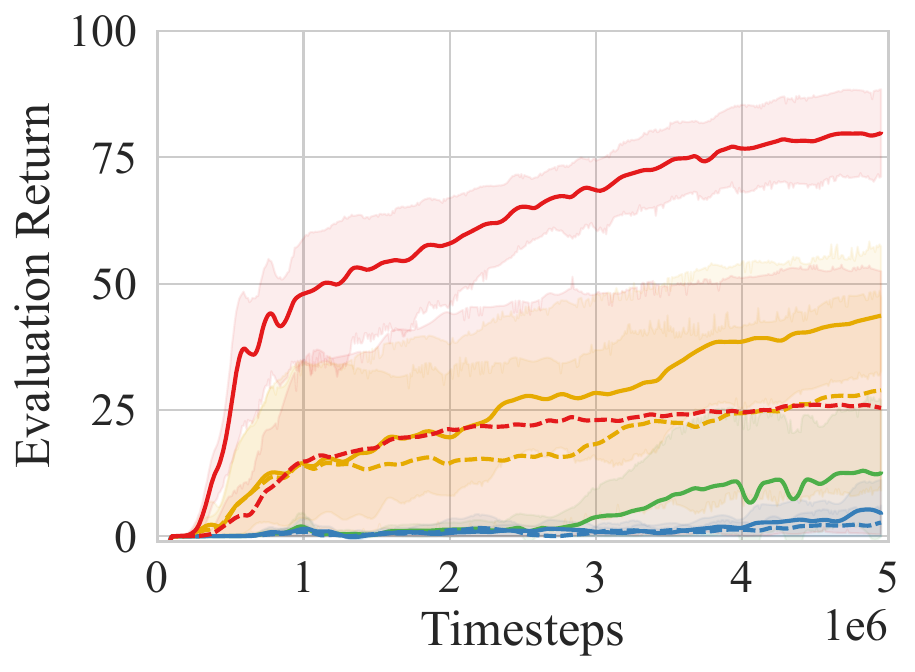}
    \caption{AdversarialPursuit w/ 45 agents}
\end{subfigure}
\begin{subfigure}[t]{.32\textwidth}
    \centering
    \includegraphics[width=\textwidth]{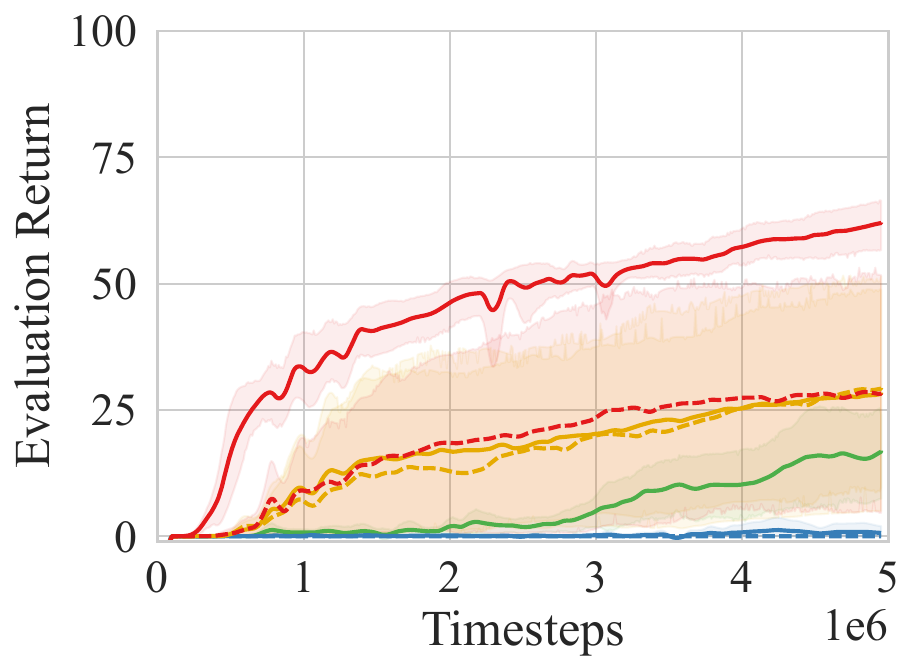}
    \caption{AdversarialPursuit w/ 61 agents}
\end{subfigure}
\begin{subfigure}[t]{.32\textwidth}
    \centering
    \includegraphics[width=\textwidth]{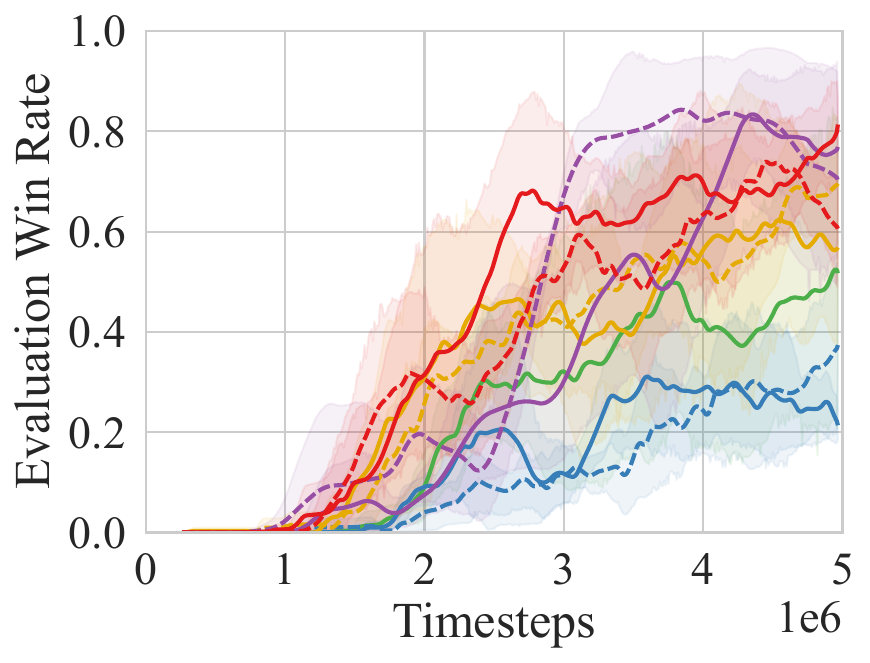}
    \caption{Battle w/ 20 agents}
\end{subfigure}
\begin{subfigure}[t]{.32\textwidth}
    \centering
    \includegraphics[width=\textwidth]{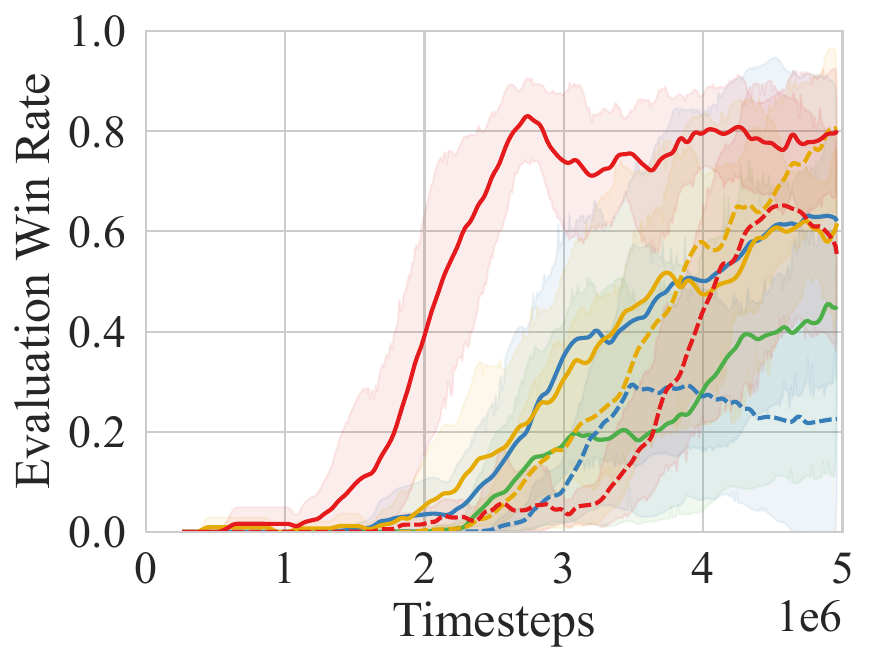}
    \caption{Battle w/ 42 agents}
\end{subfigure}
\begin{subfigure}[t]{.32\textwidth}
    \centering
    \includegraphics[width=\textwidth]{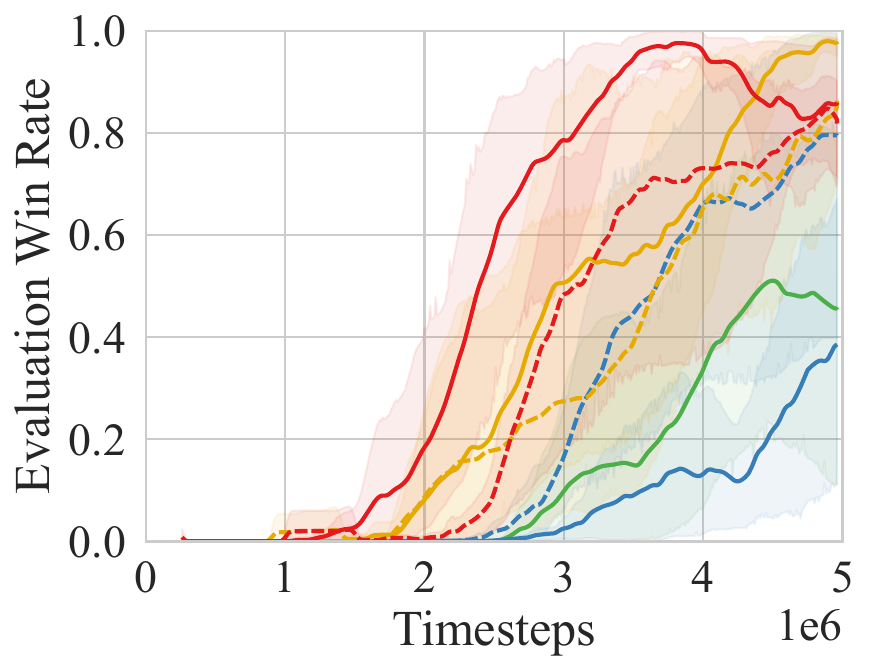}
    \caption{Battle w/ 64 agents}
\end{subfigure}
\caption{Performance comparison with baselines on MAgent tasks. Solid lines represent communication budgets of $K = 1$, while dashed lines represent budgets of $K = \lceil\log_2N\rceil$. Runs requiring more than 40 GB of GPU memory are excluded due to extreme training costs compared to other methods. \looseness=-1}
\label{fig:results_magent}
\vskip -0.25in
\end{figure}

\begin{table}[t]
\centering
\begin{threeparttable}
\caption{Performance comparison with baselines on IMP tasks. Results are reported as the mean and standard deviation of the percentage of normalized discounted rewards relative to expert-based heuristic policies, following \citet{benchmark_IMP}, with details in \cref{supp: env_details}. The best-performing method is indicated in \textbf{bold}, and the second best is \underline{underlined}.}
\label{tab:results_imp}
\begin{tabular}{lccccc}
\toprule
\multirow{2}{*}{Scenario} & \multicolumn{1}{c}{QMIX} & \multicolumn{2}{c}{ER} & \multicolumn{2}{c}{ExpoComm} \\
\cmidrule(lr){2-2} \cmidrule(lr){3-4} \cmidrule(lr){5-6}
& $K=0$ & $K=1$ & $K=\lceil\log_2N\rceil$ & $K=1$ & $K=\lceil\log_2N\rceil$ \\
\midrule
\multicolumn{6}{c}{$N=50$} \\
\midrule
Uncorrelated   & $26.42\,\scalebox{0.8}{($3.43$)}$     & $24.91\,\scalebox{0.8}{($3.77$)}$  & $26.62\,\scalebox{0.8}{($2.03$)}$  & $\underline{27.31\,\scalebox{0.8}{($2.26$)}}$  & $\mathbf{28.26\,\scalebox{0.8}{($2.51$)}}$ \\
Correlated     & $24.81\,\scalebox{0.8}{($4.16$)}$     & $34.63\,\scalebox{0.8}{($9.72$)}$  & $34.76\,\scalebox{0.8}{($5.07$)}$  & $\mathbf{43.82\,\scalebox{0.8}{($6.33$)}}$  & $\underline{40.01\,\scalebox{0.8}{($3.19$)}}$ \\
OWF            & $62.45\,\scalebox{0.8}{($3.46$)}$     & $62.99\,\scalebox{0.8}{($3.02$)}$  & $61.70\,\scalebox{0.8}{($4.62$)}$  & $\underline{64.66\,\scalebox{0.8}{($0.26$)}}$  & $\mathbf{65.19\,\scalebox{0.8}{($0.51$)}}$ \\
\midrule
\multicolumn{6}{c}{$N=100$} \\
\midrule
Uncorrelated  & $12.86\,\scalebox{0.8}{($6.88$)}$     & $21.94\,\scalebox{0.8}{($5.97$)}$  & $18.36\,\scalebox{0.8}{($12.92$)}$ & $\underline{27.34\,\scalebox{0.8}{($13.32$)}}$ & $\mathbf{27.81\,\scalebox{0.8}{($5.71$)}}$ \\
Correlated    & $-40.20\,\scalebox{0.8}{($96.35$)}$ & $-65.14\,\scalebox{0.8}{($65.08$)}$ & $9.84\,\scalebox{0.8}{($32.27$)}$ & $\mathbf{19.17\,\scalebox{0.8}{($23.94$)}}$ & $\underline{17.25\,\scalebox{0.8}{($22.70$)}}$ \\
OWF           & $65.55\,\scalebox{0.8}{($0.53$)}$     & $\mathbf{66.70\,\scalebox{0.8}{($0.50$)}}$    & $65.92\,\scalebox{0.8}{($0.87$)}$    & $65.26\,\scalebox{0.8}{($1.34$)}$    & $\underline{66.23\,\scalebox{0.8}{($0.38$)}}$ \\
\bottomrule
\end{tabular}
\begin{tablenotes}\footnotesize
\item[1] DGN+TarMAC is not suitable for this benchmark because it requires the physical positions of agents, which are not available in this environment.
\item[2] Methods that require more than 40 GB GPU memory are excluded from comparison due to the extreme training cost compared to other methods.
\end{tablenotes}
\end{threeparttable}
\end{table}

\paragraph{Benchmark results} We present the comparative performance of ExpoComm and baselines in MAgent and IMP environments with \cref{fig:results_magent} and \cref{tab:results_imp}, respectively. Overall, ExpoComm demonstrates superior performance in these large-scale benchmarks under both communication budgets, underscoring the scalability and robustness of ExpoComm strategies. 
Notably, the one-peer version of ExpoComm achieves the best performance in most scenarios, despite communication costs that only grow linearly with the number of agents. This makes it the most suitable method for handling large-scale MARL communication problems under very low communication budgets. Additional visualization results to illustrate the learned policies are provided in \cref{supp: vis_results}.

\begin{figure}
    \centering
    \includegraphics[width=\linewidth]{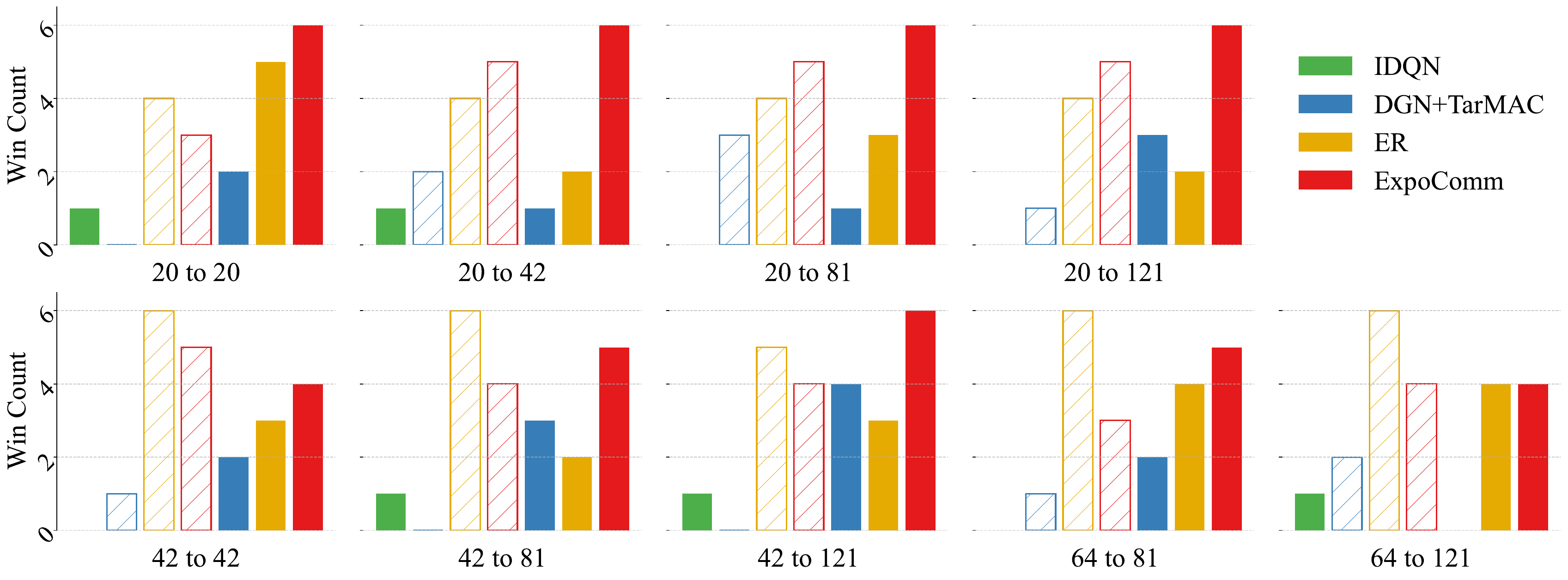}
    \caption{Zero-shot transfer results on \texttt{Battle} scenario. The subtitle ``X to Y'' indicates that methods are trained with X agents and tested with Y agents. Filled bars represent communication budgets of $K = \lceil\log_2N\rceil$, while hatched bars represent budgets of $K = 1$. Baseline CommFormer is not excluded in this experiment because it learns a fixed peer-to-peer communication topology among agents in a specific scenario and it is non-trivial to transfer such topology to scenarios with different numbers of agents. \looseness-1}
    \label{fig:results_transfer}
    \vskip -0.2in
\end{figure}

\paragraph{Zero-shot transfer}
Similar to the experimental settings suggested by \citet{ToM2C}, we test the zero-shot transfer ability of our proposed ExpoComm and the baseline methods, reporting the results in \cref{fig:results_transfer}. Specifically, we train the agent policies and their corresponding communication policies in scenarios with smaller numbers of agents and directly test these policies against each other in larger agent scenarios in the competitive task \texttt{Battle}. We test each pair of methods over 200 games, record the method with more wins as the winner, and summarize the results in \cref{fig:results_transfer}. We observe that both ER and ExpoComm demonstrate good transfer ability compared to other baselines, with ExpoComm performing better under smaller communication budgets. The superior transfer ability of ER and ExpoComm may be attributed to the grounding of messages, which reflects global information.

\begin{figure}[t]
\centering
% \subfigure[]{
% \includegraphics[width=0.7\textwidth]{Styles/figs/results/0515_legend_cropped.pdf}
%     }
\raisebox{-\height}{\includegraphics[width=0.7\textwidth]{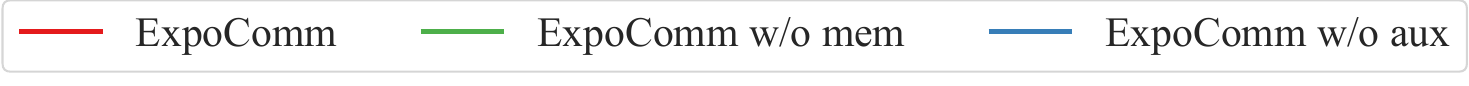}}
\par
\begin{subfigure}[t]{.32\textwidth}
    \centering
    \includegraphics[width=\textwidth]{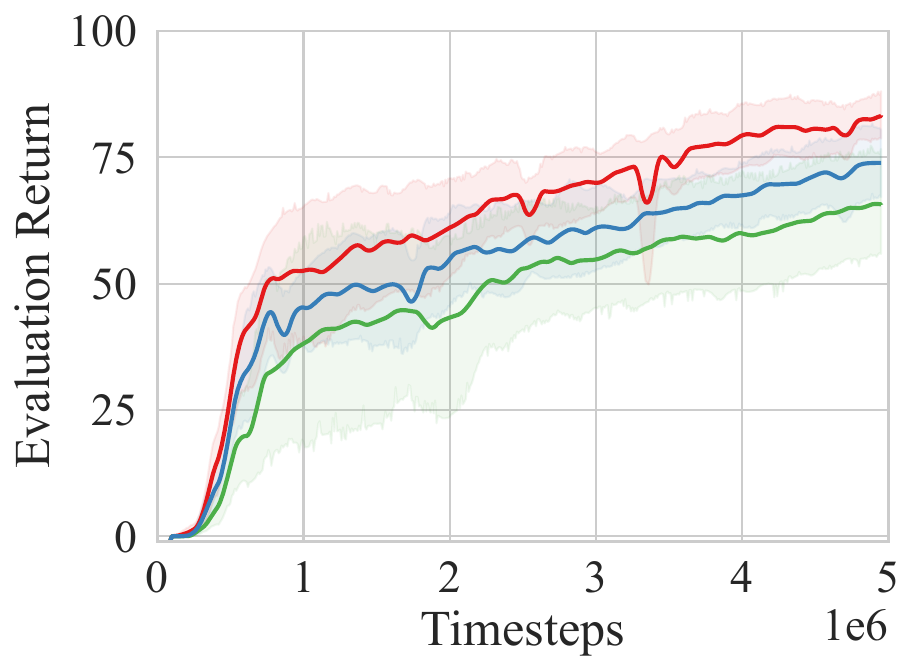}
    \caption{AdversarialPursuit w/ 25 agents}
\end{subfigure}
\begin{subfigure}[t]{.32\textwidth}
    \centering
    \includegraphics[width=\textwidth]{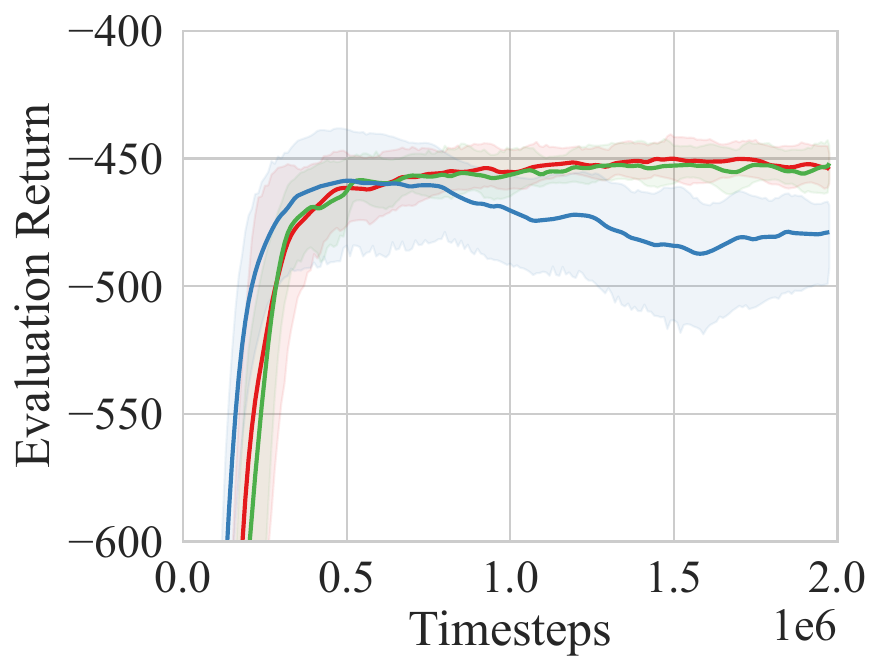}
    \caption{OWF w/ 50 agents}
\end{subfigure}
\begin{subfigure}[t]{.32\textwidth}
    \centering
    \includegraphics[width=\textwidth]{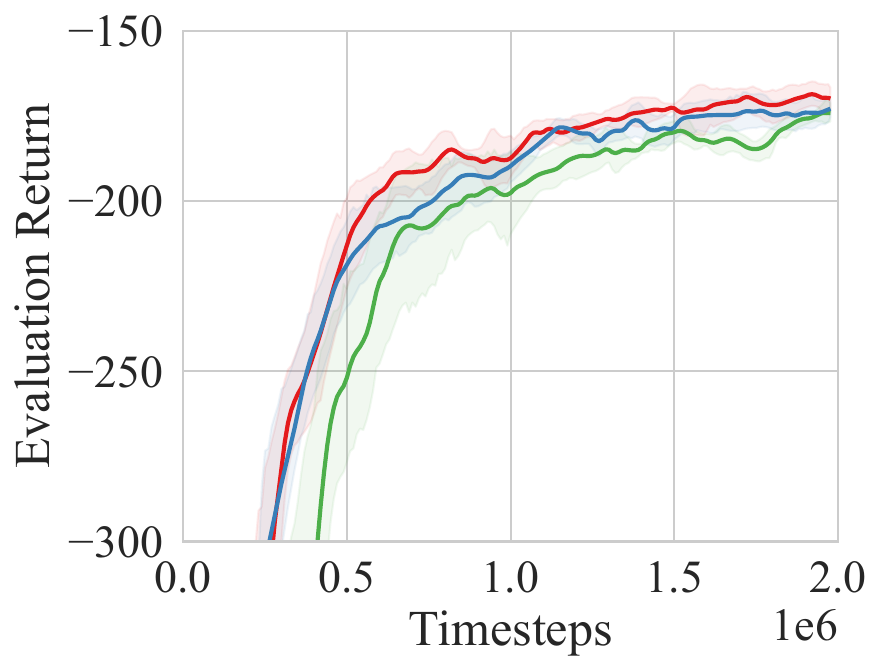}
    \caption{Uncorrelated w/ 50 agents}
\end{subfigure}

\caption{Ablation studies on MAgent and IMP benchmarks.}
\label{fig:results_ablation}
\vskip -0.2in
\end{figure}

\paragraph{Ablation studies}
We conduct ablation studies to assess the impact of various design elements in ExpoComm, with results presented in \cref{fig:results_ablation}. In particular, we compare ExpoComm with two ablations: (i) \textit{ExpoComm w/o mem}, in which the message generators are not memory-based as described in \cref{sec: network_design}. (ii) \textit{ExpoComm w/o aux}, which lacks the auxiliary loss term described in \cref{sec: training_details}. From the results, we see that removing the memory blocks from the message generators hinders effective message generation, especially in scenarios with strong time correlation such as \texttt{AdversarialPursuit}. Auxiliary tasks primarily aid in grounding the messages, without which the messages could lack guidance and even be detrimental to decision-making.

\paragraph{Discussion}
As discussed in \cref{sec: related_work}, existing methods suitable for large-scale multi-agent communication fall into two categories: position-based methods (e.g., DGN+TarMAC) and GNN-based methods (e.g., CommFormer). We propose a third category based on specific graph structures. In this category, ExpoComm utilizes exponential topologies, and we construct the baseline ER using Erdős–Rényi random graphs. Based on our analysis and experiments, the advantages of ExpoComm are as follows:
\begin{itemize}
    \item \textbf{Superior task performance}: Due to the small diameter of exponential graphs and memory-based message generators, ExpoComm facilitates fast message dissemination among all agents. It efficiently collects and carries local information from all agents, aiding decentralized decision-making. This advantage is supported by experiments shown in \cref{fig:results_magent} and \cref{tab:results_imp}.
    \item \textbf{Low communication costs}: The compact size of exponential graphs ensures that ExpoComm's communication costs scale (near-)linearly with the number of agents $N$, crucial for managing communication costs in multi-agent systems. Unlike position-based methods or ER, which can only control the number of in-edges or out-edges without global scheduling, ExpoComm naturally balances communication overhead across agents.
    \item \textbf{Versatile adaptability}: ExpoComm shows strong transferability across different numbers of agents, as seen in \cref{fig:results_transfer}. This is due to its global message dissemination strategies, which focus on overall communication rather than pairwise relationships, allowing it to adapt to more agents. Additionally, ExpoComm handles a wide range of tasks, regardless of the task nature or agent count. Unlike position-based methods, which may struggle with non-gridworld tasks like IMP due to assumptions about knowledge of agent locations, ExpoComm makes no such assumptions. Moreover, while learning effective pairwise communication topologies using GNNs can lead to significant GPU memory consumption, ExpoComm bypasses these challenges. It does not rely on the expensive task of learning a scenario-specific communication topology guided by the MARL task itself but instead uses a well-designed rule-based topology based on the communication desiderata analyzed in \cref{sec: comm_graph_desiderata}.

\end{itemize}

% why others fail
% the advantages of ExpoComm

\section{Conclusions}
% \textcolor{blue}{0.5 page}

In this work, we explored scalable communication strategies in MARL and introduced ExpoComm, an exponential topology-enabled communication protocol. We proposed a framework with communication topologies featuring small diameters for fast information dissemination and small graph sizes for low communication overhead. This framework is complemented by memory-based message processors and message grounding through auxiliary objectives to achieve effective global information representation. Despite requiring only (near-)linear communication costs relative to the number of agents, ExpoComm demonstrated superior performance and strong transferability on large-scale benchmarks like MAgent and IMP. This study highlights the potential for enhancing the scalability of MARL communication strategies through the explicit design of communication topologies. \looseness=-1

\newpage

\section*{Reproducibility Statement}
Method and implementation details are provided in \cref{sec: method}, \cref{supp: net_arch_hyperparams}, and \cref{supp: implement_details}. Experiment settings and details are described in \cref{sec: experiment_setups} and \cref{supp: env_details}. Information about the experimental infrastructure is available in \cref{supp: infrast}. The
code is publicly available at \url{https://github.com/LXXXXR/ExpoComm}.
% The code will be released publicly after the review process.

\section*{Acknowledgment}
This work was supported by the Hong Kong Research Grants Council under the NSFC/RGC Collaborative Research Scheme grant CRS\_HKUST603/22 and the Shanghai Sailing Program 24YF2710200. We thank the anonymous reviewers for their valuable feedback and suggestions.

\bibliography{ref}

\begin{thebibliography}{57}
\providecommand{\natexlab}[1]{#1}
\providecommand{\url}[1]{\texttt{#1}}
\expandafter\ifx\csname urlstyle\endcsname\relax
  \providecommand{\doi}[1]{doi: #1}\else
  \providecommand{\doi}{doi: \begingroup \urlstyle{rm}\Url}\fi

\bibitem[Ackermann et~al.(2019)Ackermann, Gabler, Osa, and Sugiyama]{MATD3}
Johannes Ackermann, Volker Gabler, Takayuki Osa, and Masashi Sugiyama.
\newblock Reducing overestimation bias in multi-agent domains using double centralized critics.
\newblock \emph{arXiv preprint arXiv:1910.01465}, 2019.

\bibitem[Assran et~al.(2019)Assran, Loizou, Ballas, and Rabbat]{exp_graph_decentralized_learning_first}
Mahmoud Assran, Nicolas Loizou, Nicolas Ballas, and Mike Rabbat.
\newblock Stochastic gradient push for distributed deep learning.
\newblock In \emph{Proceedings of the 36th International Conference on Machine Learning}, pp.\  344--353. PMLR, 2019.

\bibitem[Chu et~al.(2020)Chu, Chinchali, and Katti]{NeurComm}
Tianshu Chu, Sandeep Chinchali, and Sachin Katti.
\newblock Multi-agent reinforcement learning for networked system control.
\newblock In \emph{Proceedings of the 8th International Conference on Learning Representations}, 2020.

\bibitem[Cui et~al.(2022)Cui, Tahir, Ekinci, Elshamanhory, Eich, Li, and Koeppl]{population_survey}
Kai Cui, Anam Tahir, Gizem Ekinci, Ahmed Elshamanhory, Yannick Eich, Mengguang Li, and Heinz Koeppl.
\newblock A survey on large-population systems and scalable multi-agent reinforcement learning.
\newblock \emph{arXiv preprint arXiv:2209.03859}, 2022.

\bibitem[Das et~al.(2019)Das, Gervet, Romoff, Batra, Parikh, Rabbat, and Pineau]{TarMAC}
Abhishek Das, Th{\'e}ophile Gervet, Joshua Romoff, Dhruv Batra, Devi Parikh, Mike Rabbat, and Joelle Pineau.
\newblock Tarmac: Targeted multi-agent communication.
\newblock In \emph{Proceedings of the 36th International Conference on Machine Learning}, pp.\  1538--1546. PMLR, 2019.

\bibitem[Ding et~al.(2020)Ding, Huang, and Lu]{i2c}
Ziluo Ding, Tiejun Huang, and Zongqing Lu.
\newblock Learning individually inferred communication for multi-agent cooperation.
\newblock In \emph{Advances in Neural Information Processing Systems}, volume~33, pp.\  22069--22079, 2020.

\bibitem[Erdos et~al.(1960)Erdos, R{\'e}nyi, et~al.]{ER_graph}
Paul Erdos, Alfr{\'e}d R{\'e}nyi, et~al.
\newblock On the evolution of random graphs.
\newblock \emph{Publ. math. inst. hung. acad. sci}, 5\penalty0 (1):\penalty0 17--60, 1960.

\bibitem[Foerster et~al.(2016)Foerster, Assael, De~Freitas, and Whiteson]{DIAL_RIAL}
Jakob Foerster, Ioannis~Alexandros Assael, Nando De~Freitas, and Shimon Whiteson.
\newblock Learning to communicate with deep multi-agent reinforcement learning.
\newblock In \emph{Advances in Neural Information Processing Systems}, volume~29, 2016.

\bibitem[Foerster et~al.(2018)Foerster, Farquhar, Afouras, Nardelli, and Whiteson]{COMA}
Jakob Foerster, Gregory Farquhar, Triantafyllos Afouras, Nantas Nardelli, and Shimon Whiteson.
\newblock Counterfactual multi-agent policy gradients.
\newblock In \emph{Proceedings of the AAAI Conference on Artificial Intelligence}, volume~32, 2018.

\bibitem[Freed et~al.(2020)Freed, Sartoretti, Hu, and Choset]{noisy_channel}
Benjamin Freed, Guillaume Sartoretti, Jiaheng Hu, and Howie Choset.
\newblock Communication learning via backpropagation in discrete channels with unknown noise.
\newblock In \emph{Proceedings of the AAAI conference on artificial intelligence}, volume~34, pp.\  7160--7168, 2020.

\bibitem[Guan et~al.(2022)Guan, Chen, Yuan, Wang, Yin, Zhang, and Yu]{MASIA}
Cong Guan, Feng Chen, Lei Yuan, Chenghe Wang, Hao Yin, Zongzhang Zhang, and Yang Yu.
\newblock Efficient multi-agent communication via self-supervised information aggregation.
\newblock In \emph{Advances in Neural Information Processing Systems}, volume~35, pp.\  1020--1033, 2022.

\bibitem[Hu et~al.(2021)Hu, Zhu, Zhao, Zhao, and Hao]{ETC}
Guangzheng Hu, Yuanheng Zhu, Dongbin Zhao, Mengchen Zhao, and Jianye Hao.
\newblock Event-triggered communication network with limited-bandwidth constraint for multi-agent reinforcement learning.
\newblock \emph{IEEE Transactions on Neural Networks and Learning Systems}, 2021.

\bibitem[Hu et~al.(2024)Hu, Shen, Zhang, and Tao]{Commformer}
Shengchao Hu, Li~Shen, Ya~Zhang, and Dacheng Tao.
\newblock Learning multi-agent communication from graph modeling perspective.
\newblock In \emph{Proceedings of the 12th International Conference on Learning Representations}, 2024.

\bibitem[Jiang \& Lu(2018)Jiang and Lu]{atoc}
Jiechuan Jiang and Zongqing Lu.
\newblock Learning attentional communication for multi-agent cooperation.
\newblock In \emph{Advances in Neural Information Processing Systems}, volume~31, 2018.

\bibitem[Jiang et~al.(2020)Jiang, Dun, Huang, and Lu]{DGN}
Jiechuan Jiang, Chen Dun, Tiejun Huang, and Zongqing Lu.
\newblock Graph convolutional reinforcement learning.
\newblock In \emph{Proceedings of the 8th International Conference on Learning Representations}, 2020.

\bibitem[Kim et~al.(2019)Kim, Moon, Hostallero, Kang, Lee, Son, and Yi]{schedNet}
Daewoo Kim, Sangwoo Moon, David Hostallero, Wan~Ju Kang, Taeyoung Lee, Kyunghwan Son, and Yung Yi.
\newblock Learning to schedule communication in multi-agent reinforcement learning.
\newblock In \emph{Proceedings of the 7th International Conference on Learning Representations}, 2019.

\bibitem[Kong et~al.(2021)Kong, Lin, Koloskova, Jaggi, and Stich]{expG_consensus}
Lingjing Kong, Tao Lin, Anastasia Koloskova, Martin Jaggi, and Sebastian Stich.
\newblock Consensus control for decentralized deep learning.
\newblock In \emph{Proceedings of the 38th International Conference on Machine Learning}, pp.\  5686--5696, 2021.

\bibitem[Kraemer \& Banerjee(2016)Kraemer and Banerjee]{ctde_kraemer2016multi}
Landon Kraemer and Bikramjit Banerjee.
\newblock Multi-agent reinforcement learning as a rehearsal for decentralized planning.
\newblock \emph{Neurocomputing}, 190:\penalty0 82--94, 2016.

\bibitem[Leroy et~al.(2024)Leroy, Morato, Pisane, Kolios, and Ernst]{benchmark_IMP}
Pascal Leroy, Pablo~G Morato, Jonathan Pisane, Athanasios Kolios, and Damien Ernst.
\newblock {IMP-MARL}: a suite of environments for large-scale infrastructure management planning via {MARL}.
\newblock In \emph{Advances in Neural Information Processing Systems}, volume~36, 2024.

\bibitem[Li \& Zhang(2024)Li and Zhang]{comm_aux_loss}
Xinran Li and Jun Zhang.
\newblock Context-aware communication for multi-agent reinforcement learning.
\newblock In \emph{Proceedings of the 23rd International Conference on Autonomous Agents and Multiagent Systems}, pp.\  1156--1164, 2024.

\bibitem[Liu et~al.(2020)Liu, Wang, Hu, Hao, Chen, and Gao]{G2ANet}
Yong Liu, Weixun Wang, Yujing Hu, Jianye Hao, Xingguo Chen, and Yang Gao.
\newblock Multi-agent game abstraction via graph attention neural network.
\newblock In \emph{Proceedings of the AAAI conference on artificial intelligence}, volume~34, pp.\  7211--7218, 2020.

\bibitem[Lo et~al.(2024)Lo, Sengupta, Foerster, and Noukhovitch]{CACL}
Yat~Long Lo, Biswa Sengupta, Jakob~Nicolaus Foerster, and Michael Noukhovitch.
\newblock Learning multi-agent communication with contrastive learning.
\newblock In \emph{Proceedings of the 12th International Conference on Learning Representations}, 2024.

\bibitem[Lowe et~al.(2017)Lowe, Wu, Tamar, Harb, Abbeel, and Mordatch]{MADDPG}
Ryan Lowe, Yi~Wu, Aviv Tamar, Jean Harb, Pieter Abbeel, and Igor Mordatch.
\newblock Multi-agent actor-critic for mixed cooperative-competitive environments.
\newblock In \emph{Advances in Neural Information Processing Systems}, volume~30, 2017.

\bibitem[Lyu et~al.(2021)Lyu, Xiao, Daley, and Amato]{ctde_lyu2021contrasting}
Xueguang Lyu, Yuchen Xiao, Brett Daley, and Christopher Amato.
\newblock Contrasting centralized and decentralized critics in multi-agent reinforcement learning.
\newblock In \emph{Proceedings of the 20th International Conference on Autonomous Agents and Multiagent Systems}, pp.\  844--852, 2021.

\bibitem[Ma et~al.(2024)Ma, Li, Du, Dong, and Yang]{large_scale_networked_MARL}
Chengdong Ma, Aming Li, Yali Du, Hao Dong, and Yaodong Yang.
\newblock Efficient and scalable reinforcement learning for large-scale network control.
\newblock \emph{Nature Machine Intelligence}, pp.\  1--15, 2024.

\bibitem[Niu et~al.(2021)Niu, Paleja, and Gombolay]{MAGIC}
Yaru Niu, Rohan~R Paleja, and Matthew~C Gombolay.
\newblock Multi-agent graph-attention communication and teaming.
\newblock In \emph{Proceedings of the 20th International Conference on Autonomous Agents and MultiAgent Systems}, volume~21, pp.\  964--97, 2021.

\bibitem[Oliehoek \& Amato(2016)Oliehoek and Amato]{pomdp_oliehoek2016concise}
Frans~A Oliehoek and Christopher Amato.
\newblock \emph{A concise introduction to decentralized POMDPs}.
\newblock Springer, 2016.

\bibitem[Oord et~al.(2018)Oord, Li, and Vinyals]{InfoNCE}
Aaron van~den Oord, Yazhe Li, and Oriol Vinyals.
\newblock Representation learning with contrastive predictive coding.
\newblock \emph{arXiv preprint arXiv:1807.03748}, 2018.

\bibitem[Papoudakis et~al.(2021)Papoudakis, Christianos, Schäfer, and Albrecht]{epymarl}
Georgios Papoudakis, Filippos Christianos, Lukas Schäfer, and Stefano~V. Albrecht.
\newblock Benchmarking multi-agent deep reinforcement learning algorithms in cooperative tasks.
\newblock In \emph{Proceedings of the Neural Information Processing Systems Track on Datasets and Benchmarks}, 2021.
\newblock URL \url{http://arxiv.org/abs/2006.07869}.

\bibitem[Peng et~al.(2021)Peng, Rashid, Schroeder~de Witt, Kamienny, Torr, B{\"o}hmer, and Whiteson]{facmac_mamujoco}
Bei Peng, Tabish Rashid, Christian Schroeder~de Witt, Pierre-Alexandre Kamienny, Philip Torr, Wendelin B{\"o}hmer, and Shimon Whiteson.
\newblock Facmac: Factored multi-agent centralised policy gradients.
\newblock In \emph{Advances in Neural Information Processing Systems}, volume~34, pp.\  12208--12221, 2021.

\bibitem[Peng et~al.(2017)Peng, Wen, Yang, Yuan, Tang, Long, and Wang]{BiCNet}
Peng Peng, Ying Wen, Yaodong Yang, Quan Yuan, Zhenkun Tang, Haitao Long, and Jun Wang.
\newblock Multiagent bidirectionally-coordinated nets: Emergence of human-level coordination in learning to play starcraft combat games.
\newblock \emph{arXiv preprint arXiv:1703.10069}, 2017.

\bibitem[Rashid et~al.(2020)Rashid, Samvelyan, De~Witt, Farquhar, Foerster, and Whiteson]{QMIX}
Tabish Rashid, Mikayel Samvelyan, Christian~Schroeder De~Witt, Gregory Farquhar, Jakob Foerster, and Shimon Whiteson.
\newblock Monotonic value function factorisation for deep multi-agent reinforcement learning.
\newblock \emph{The Journal of Machine Learning Research}, 21\penalty0 (1):\penalty0 7234--7284, 2020.

\bibitem[Samvelyan et~al.(2019)Samvelyan, Rashid, Schroeder~de Witt, Farquhar, Nardelli, Rudner, Hung, Torr, Foerster, and Whiteson]{smac}
Mikayel Samvelyan, Tabish Rashid, Christian Schroeder~de Witt, Gregory Farquhar, Nantas Nardelli, Tim~GJ Rudner, Chia-Man Hung, Philip~HS Torr, Jakob Foerster, and Shimon Whiteson.
\newblock The starcraft multi-agent challenge.
\newblock In \emph{Proceedings of the 18th International Conference on Autonomous Agents and MultiAgent Systems}, pp.\  2186--2188, 2019.

\bibitem[Schmidt et~al.(2022)Schmidt, Brosig, Plinge, Eskofier, and Mutschler]{traffic_scalibility}
Lukas~M Schmidt, Johanna Brosig, Axel Plinge, Bjoern~M Eskofier, and Christopher Mutschler.
\newblock An introduction to multi-agent reinforcement learning and review of its application to autonomous mobility.
\newblock In \emph{IEEE 25th International Conference on Intelligent Transportation Systems (ITSC)}, pp.\  1342--1349. IEEE, 2022.

\bibitem[Seuken \& Zilberstein(2007)Seuken and Zilberstein]{marl_delivery}
Sven Seuken and Shlomo Zilberstein.
\newblock Improved memory-bounded dynamic programming for decentralized pomdps.
\newblock In \emph{Proceedings of the Twenty-Third Conference on Uncertainty in Artificial Intelligence}, pp.\  344--351, 2007.

\bibitem[Singh et~al.(2019)Singh, Jain, and Sukhbaatar]{ic3}
Amanpreet Singh, Tushar Jain, and Sainbayar Sukhbaatar.
\newblock Learning when to communicate at scale in multiagent cooperative and competitive tasks.
\newblock In \emph{Proceedings of the 7th International Conference on Learning Representations}, 2019.

\bibitem[Sukhbaatar et~al.(2016)Sukhbaatar, Fergus, et~al.]{CommNet}
Sainbayar Sukhbaatar, Rob Fergus, et~al.
\newblock Learning multiagent communication with backpropagation.
\newblock In \emph{Advances in Neural Information Processing Systems}, volume~29, 2016.

\bibitem[Swamy et~al.(2020)Swamy, Reddy, Levine, and Dragan]{marl_robot_swamy2020scaled}
Gokul Swamy, Siddharth Reddy, Sergey Levine, and Anca~D Dragan.
\newblock Scaled autonomy: Enabling human operators to control robot fleets.
\newblock In \emph{2020 IEEE International Conference on Robotics and Automation}, pp.\  5942--5948. IEEE, 2020.

\bibitem[Terry et~al.(2020)Terry, Black, and Jayakumar]{magent2}
Jordan~K Terry, Benjamin Black, and Mario Jayakumar.
\newblock Magent.
\newblock \url{https://github.com/Farama-Foundation/MAgent}, 2020.
\newblock GitHub repository.

\bibitem[Wang et~al.(2020{\natexlab{a}})Wang, Tantia, Ballas, and Rabbat]{SlowMo}
Jianyu Wang, Vinayak Tantia, Nicolas Ballas, and Michael~G. Rabbat.
\newblock Slowmo: Improving communication-efficient distributed {SGD} with slow momentum.
\newblock In \emph{Proceedings of 8th International Conference on Learning Representations}, 2020{\natexlab{a}}.

\bibitem[Wang et~al.(2020{\natexlab{b}})Wang, Wang, Zheng, and Zhang]{NDQ}
Tonghan Wang, Jianhao Wang, Chongyi Zheng, and Chongjie Zhang.
\newblock Learning nearly decomposable value functions via communication minimization.
\newblock In \emph{Proceedings of the International Conference on Learning Representations}, 2020{\natexlab{b}}.

\bibitem[Wang et~al.(2015)Wang, Gu, Yang, Wang, and Hao]{wang2014rpnoc}
Xiaolu Wang, Huaxi Gu, Yintang Yang, Kun Wang, and Qinfen Hao.
\newblock {RPNoC}: A ring-based packet-switched optical network-on-chip.
\newblock \emph{IEEE Photonics Technology Letters}, 27\penalty0 (4):\penalty0 423--426, 2015.

\bibitem[Wang et~al.(2016)Wang, Gu, Yang, Wang, and Hao]{ExpG_chip_design}
Xiaolu Wang, Huaxi Gu, Yintang Yang, Kun Wang, and Qinfen Hao.
\newblock A highly scalable optical network-on-chip with small network diameter and deadlock freedom.
\newblock \emph{IEEE Transactions on Very Large Scale Integration (VLSI) Systems}, 24\penalty0 (12):\penalty0 3424--3436, 2016.

\bibitem[Wang et~al.(2022)Wang, Zhong, Xu, and Wang]{ToM2C}
Yuanfei Wang, Fangwei Zhong, Jing Xu, and Yizhou Wang.
\newblock {ToM2C}: Target-oriented multi-agent communication and cooperation with theory of mind.
\newblock In \emph{Proceedings of the 10th International Conference on Learning Representations}, 2022.

\bibitem[Weil et~al.(2024)Weil, Bao, Abboud, and Meuser]{neighbor_graph}
Jannis Weil, Zhenghua Bao, Osama Abboud, and Tobias Meuser.
\newblock Towards generalizability of multi-agent reinforcement learning in graphs with recurrent message passing.
\newblock In \emph{Proceedings of the 23rd International Conference on Autonomous Agents and Multiagent Systems}, pp.\  1919--1927, 2024.

\bibitem[Yang et~al.(2023)Yang, Liu, Jiang, Zhang, Zhao, Song, and Bian]{inventory_scalibility}
Xianliang Yang, Zhihao Liu, Wei Jiang, Chuheng Zhang, Li~Zhao, Lei Song, and Jiang Bian.
\newblock A versatile multi-agent reinforcement learning benchmark for inventory management.
\newblock \emph{arXiv preprint arXiv:2306.07542}, 2023.

\bibitem[Ying et~al.(2021)Ying, Yuan, Chen, Hu, Pan, and Yin]{exp_graph_decentralized_exact_avg}
Bicheng Ying, Kun Yuan, Yiming Chen, Hanbin Hu, Pan Pan, and Wotao Yin.
\newblock Exponential graph is provably efficient for decentralized deep training.
\newblock In \emph{Advances in Neural Information Processing Systems}, volume~34, pp.\  13975--13987, 2021.

\bibitem[Ying \& Dayong(2005)Ying and Dayong]{marl_dis}
Wang Ying and Sang Dayong.
\newblock Multi-agent framework for third party logistics in {E}-commerce.
\newblock \emph{Expert Systems with Applications}, 29\penalty0 (2):\penalty0 431--436, 2005.

\bibitem[Yu et~al.(2022)Yu, Velu, Vinitsky, Gao, Wang, Bayen, and Wu]{MAPPO}
Chao Yu, Akash Velu, Eugene Vinitsky, Jiaxuan Gao, Yu~Wang, Alexandre Bayen, and Yi~Wu.
\newblock The surprising effectiveness of ppo in cooperative multi-agent games.
\newblock In \emph{Advances in Neural Information Processing Systems}, volume~35, pp.\  24611--24624, 2022.

\bibitem[Yuan et~al.(2021)Yuan, Chen, Huang, Zhang, Pan, Xu, and Yin]{ExpG_largebs}
Kun Yuan, Yiming Chen, Xinmeng Huang, Yingya Zhang, Pan Pan, Yinghui Xu, and Wotao Yin.
\newblock Decentlam: Decentralized momentum sgd for large-batch deep training.
\newblock In \emph{Proceedings of the IEEE/CVF International Conference on Computer Vision}, pp.\  3029--3039, 2021.

\bibitem[Yuan et~al.(2022)Yuan, Wang, Zhang, Wang, Zhang, Yu, and Zhang]{MAIC}
Lei Yuan, Jianhao Wang, Fuxiang Zhang, Chenghe Wang, Zongzhang Zhang, Yang Yu, and Chongjie Zhang.
\newblock Multi-agent incentive communication via decentralized teammate modeling.
\newblock In \emph{Proceedings of the AAAI Conference on Artificial Intelligence}, volume~36, pp.\  9466--9474, 2022.

\bibitem[Yuan et~al.(2023)Yuan, Chung, Yuan, and Fu]{DACOM}
Tingting Yuan, Hwei-Ming Chung, Jie Yuan, and Xiaoming Fu.
\newblock Dacom: Learning delay-aware communication for multi-agent reinforcement learning.
\newblock In \emph{Proceedings of the AAAI Conference on Artificial Intelligence}, volume~37, pp.\  11763--11771, 2023.

\bibitem[Zhang et~al.(2018)Zhang, Yang, Liu, Zhang, and Basar]{networked_MA}
Kaiqing Zhang, Zhuoran Yang, Han Liu, Tong Zhang, and Tamer Basar.
\newblock Fully decentralized multi-agent reinforcement learning with networked agents.
\newblock In \emph{Proceedings of the 35th International Conference on Machine Learning}, pp.\  5872--5881. PMLR, 2018.

\bibitem[Zhang et~al.(2020)Zhang, Zhang, and Lin]{TMC}
Sai~Qian Zhang, Qi~Zhang, and Jieyu Lin.
\newblock Succinct and robust multi-agent communication with temporal message control.
\newblock In \emph{Advances in Neural Information Processing Systems}, volume~33, pp.\  17271--17282, 2020.

\bibitem[Zheng et~al.(2018)Zheng, Yang, Cai, Zhou, Zhang, Wang, and Yu]{benchmark_Magent}
Lianmin Zheng, Jiacheng Yang, Han Cai, Ming Zhou, Weinan Zhang, Jun Wang, and Yong Yu.
\newblock Magent: A many-agent reinforcement learning platform for artificial collective intelligence.
\newblock In \emph{Proceedings of the AAAI conference on artificial intelligence}, volume~32, 2018.

\bibitem[Zhou et~al.(2021)Zhou, Luo, Villella, Yang, Rusu, Miao, Zhang, Alban, Fadakar, Chen, et~al.]{MARL_autonomous_driving}
Ming Zhou, Jun Luo, Julian Villella, Yaodong Yang, David Rusu, Jiayu Miao, Weinan Zhang, Montgomery Alban, Iman Fadakar, Zheng Chen, et~al.
\newblock Smarts: An open-source scalable multi-agent {RL} training school for autonomous driving.
\newblock In \emph{Proceedings of the Conference on Robot Learning}, pp.\  264--285. PMLR, 2021.

\bibitem[Zhu et~al.(2024)Zhu, Dastani, and Wang]{MARL_comm_survey}
Changxi Zhu, Mehdi Dastani, and Shihan Wang.
\newblock A survey of multi-agent deep reinforcement learning with communication.
\newblock In Mehdi Dastani, Jaime~Sim{\~{a}}o Sichman, Natasha Alechina, and Virginia Dignum (eds.), \emph{Proceedings of the 23rd International Conference on Autonomous Agents and MultiAgent Systems}, pp.\  2845--2847, 2024.

\end{thebibliography}
\bibliographystyle{iclr2025_conference}

\appendix
% \section{Appendix}


\newpage

\section{Theoretical Analysis} \label{supp: theory}
In this section, we analyze the communication effect of exponential topologies.

\begin{theorem}
Suppose that $E^{t}_{ij}$ is defined by \cref{eq: one_peer_exp_adj}. Let $\tau = \lceil \log_2{(N-1)} \rceil$. Then, the following holds:
\begin{equation}
    E^{t}_{ij} \times_b E^{t+1}_{ij} \times_b \ldots E^{t+\tau-1}_{ij} = \mathbbm{1}\mathbbm{1}^T,
\end{equation}
where $\times_b$ denotes logical (Boolean) matrix multiplication.
\label{theorem: comm_effect}
\end{theorem}

\begin{remark}
    If the information at each agent remains valid within $\tau$ timesteps and there is no information loss during aggregation, the one-peer exponential topology ensures information exchange between any two agents in the system with $\tau$ timesteps.
\end{remark}

\begin{remark}
    Static exponential topologies exhibit a similar communication effect as described in \cref{theorem: comm_effect}. Specifically, $\forall i, j$ that $E^{t(\text{one-peer})}_{ij} = 1$, it holds that $E^{t(\text{stat})}_{ij} = 1$.
\end{remark}

\begin{proof}
Define function $Z:\mathbb{R}_+ \rightarrow \{0,1\}$ such that
    \begin{equation}
        Z(x) = \begin{cases}
    1,& x > 0,\\
    0,& x = 0.
    \end{cases}
    \end{equation}
    Then, for all $x,y,u,v\geq 0$, the following equivalence holds:
    \begin{equation}
        xy + wv = 0 \iff (Z(x) \times_b Z(y)) +_b (Z(u) \times_b Z(v)) = 0,
    \end{equation}
    where $\times_b$ denotes logical (Boolean) \texttt{And}, and $+_b$ denotes logical (Boolean) \texttt{Or}.
    \label{lemma: add_mul}
Now, consider the connection between $Z(x)$ and the structure of an all-one matrix. For a non-negative matrix $X$, it holds that $X_{ij} \in \mathbb{R}^+, \forall i, j \iff Z(X) = \mathbbm{1}\mathbbm{1}^T$. 

Therefore, by applying \cref{lemma: add_mul} to $E^{t}_{ij}  E^{t+1}_{ij} \ldots E^{t+\tau-1}_{ij} = \frac{2^\tau}{N} \mathbbm{1}\mathbbm{1}^T$~\citep{exp_graph_decentralized_exact_avg}, we have $E^{t}_{ij} \times_b E^{t+1}_{ij} \times_b \ldots E^{t+\tau-1}_{ij} = \mathbbm{1}\mathbbm{1}^T$.
\end{proof}

\color{black}

\section{Experiment Details} 
\subsection{Network architecture and hyperparameters} \label{supp: net_arch_hyperparams}

\paragraph{Codebase} 
Our implementation of ExpoComm and baseline algorithms is based on the following codebase:
\begin{itemize}
    \item EPyMARL~\citep{epymarl}: \url{https://github.com/uoe-agents/epymarl}
    \item CommFormer~\cite{Commformer}: \url{https://github.com/charleshsc/CommFormer}
    \item CommNet~\citep{CommNet}: \url{https://github.com/isp1tze/MAProj}
\end{itemize}

The code for ExpoComm is publicly available at \url{https://github.com/LXXXXR/ExpoComm}.

\paragraph{Neural network architecture}
Following previous work \cite{epymarl}, we employ deep neural networks consisting of multilayer perceptrons (MLPs) with rectified linear unit (ReLU) activation functions and gated recurrent units (GRUs) to parameterize the agent networks. 
In ExpoComm, the message memory blocks described in \cref{sec: network_design} are implemented using a single GRU or an attention block. The prediction network $f(\cdot; \phi)$ described in \cref{sec: training_details} is implemented using a two-layer MLP. 

\paragraph{Hyperparameters} 
To ensure a fair comparison, we implement our method and self-constructed baselines using the same codebase with the same set of hyperparameters, with the exception of method-specific ones and the learning rate. In general, we follow the common settings provided by \citet{epymarl} for MAgent benchmark and adopt the settings in IMP paper~\citep{benchmark_IMP} for the IMP benchmark. The common hyperparameters are listed in \cref{tab: comm_hyper}. The ExpoComm-specific hyperparameters are provided in \cref{tab: hyper_ExpoComm}. For learning rate, we search among $(0.0001, 0.0005)$ for ExpoComm and baselines. We use the value of $0.0005$ for base algorithms without communication; $0.0005$ for DGN+TarMAC in MAgent and $0.0001$ for DGN+TarMAC in IMP; $0.0001$ for ExpoComm in IMP with 50 agents and $0.0005$ for other scenarios. For CommFormer, we adopt the optimal value in its official implementation.

\begin{table}[th]
    \centering
    \caption{Common hyperparameters.}
    \label{tab: comm_hyper}
    \begin{threeparttable}
    \begin{tabular}{ccc}
    \toprule
    Hyperparameter & Benchmark & Value \\
    \midrule
    Hidden sizes & - & $64$ \\
    \multirow{2}{*}{Discount factor $\gamma$} & MAgent &  $0.99$ \\
    & IMP & $0.95$ \\
    \multirow{2}{*}{Batch size} & MAgent & $32$ \\
    & IMP & $64$ \\
    Replay buffer size & - &  $2000$ \\
    \multirow{2}{*}{Number of environment steps} & MAgent & $5 \times 10^{6}$ \\
    & IMP & $2 \times 10^{6}$ \\
    \multirow{2}{*}{Epsilon anneal steps} & MAgent & $5 \times 10^{5}$ \\
    & IMP & $5 \times 10^{3}$ \\
    \multirow{2}{*}{Test interval steps} & MAgent & $5 \times 10^{4}$ \\
    & IMP & $2.5 \times 10^{4}$ \\
    Number of test episode & - & $100$ \\
    \bottomrule
    \end{tabular}
    % \begin{tablenotes}
    %     \item[1] Here we adopt the per-scenario finetuned value for this hyperparameter as provided by HARL.
    % \end{tablenotes}
    \end{threeparttable}
\end{table}

% \begin{table}[th]
%     \centering
%     \caption{Common hyperparameters.}
%     \label{tab:comm_hyper}
%     \begin{threeparttable}
%     \begin{tabular}{lcc}
%     \toprule
%     Hyperparameter & MAgent & IMP \\
%     \midrule
%     Hidden sizes & \multicolumn{2}{c}{$64$} \\
%     Discount factor $\gamma$ & $0.99$ & $0.95$ \\
%     Batch size & $32$ & $64$ \\
%     Replay buffer size & \multicolumn{2}{c}{$2000$} \\
%     Number of environment steps & $5 \times 10^{6}$ & $2 \times 10^{6}$ \\
%     Epsilon anneal steps & $5 \times 10^{5}$ & $5 \times 10^{3}$ \\
%     Test interval steps & $5 \times 10^{4}$ & $2.5 \times 10^{4}$ \\
%     Number of test episodes & \multicolumn{2}{c}{$100$} \\
%     \bottomrule
%     \end{tabular}
%     \end{threeparttable}
% \end{table}

\begin{table}[th]
    \centering
    \caption{Hyperparameters used for ExpoComm.}
    \label{tab: hyper_ExpoComm}
    \begin{tabular}{cc}
    \toprule
    Hyperparameter & Value \\
    \midrule
    Auxillary loss coefficient $\alpha$  & $0.1$\\
    Temperature $\tau$  & $0.07$\\
    Number of negative pairs $M$  & $20$ \\
    \bottomrule
    \end{tabular}
\end{table}

\subsection{Environmental details} \label{supp: env_details}

\paragraph{Codebase} The environments used in this work are listed below with descriptions in \cref{tab: env}.
\begin{itemize}
    \item MAgent~\citep{benchmark_Magent,magent2}: \\\url{https://github.com/Farama-Foundation/MAgent2}
    \item IMP~\citep{benchmark_IMP}: \url{https://github.com/moratodpg/imp_marl}
\end{itemize}

\begin{table}[th]
    \centering
    \caption{Environments details.}
    \label{tab: env}
    \begin{threeparttable}
    \begin{tabular}{ccc}
    \toprule
    Environment &  Scenarios & Number of agents \\
    \midrule
    \multirow{2}{*}{MAgent} &  Adversarial Pursuit & $(25, 45, 61)$ \tnote{1} \\
    &  Battle & $(20, 42, 64)$ \tnote{2}\\
    % \hline \noalign{\vskip 2pt}
    \midrule
    \multirow{3}{*}{IMP} &  Uncorrelated: uncorrelated k-out-of-n; campaign cost & $(50, 100)$ \\
    &  Correlated: correlated k-out-of-n; campaign cost & $(50, 100)$ \\
    &  OWF: offshore wind farm; campaign cost & $(50, 100)$ \\
    \bottomrule
    \end{tabular}
    \begin{tablenotes}
        \item[1] The number of agents in this scenario is determined by setting the map size to $25, 35, 40$, respectively.
        \item[2] The number of agents in this scenario is determined by setting the map size to $45, 60, 70$, respectively.
    \end{tablenotes}
    \end{threeparttable}
\end{table}

\begin{figure}[t]
\centering
\begin{subfigure}[t]{.26\textwidth}
    \centering
    \includegraphics[width=\textwidth]{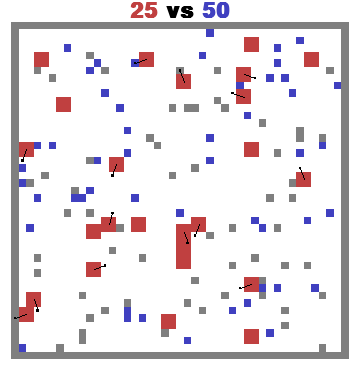}
    \caption{AdversarialPursuit.}
\end{subfigure}
\begin{subfigure}[t]{.26\textwidth}
    \centering
    \includegraphics[width=\textwidth]{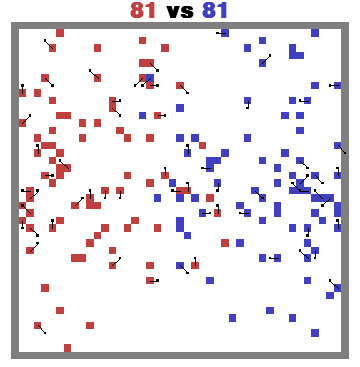}
    \caption{Battle.}
\end{subfigure}
\caption{Environments from the MAgent benchmark suite~\citep{magent2}. In each scenario, the MARL algorithms control the red agents, while the blue adversary agents are controlled by pretrained policies.}
\label{fig: Magent_env}
\end{figure}

\begin{figure}[t]
  \centering
    \centering
    \includegraphics[width=\textwidth]{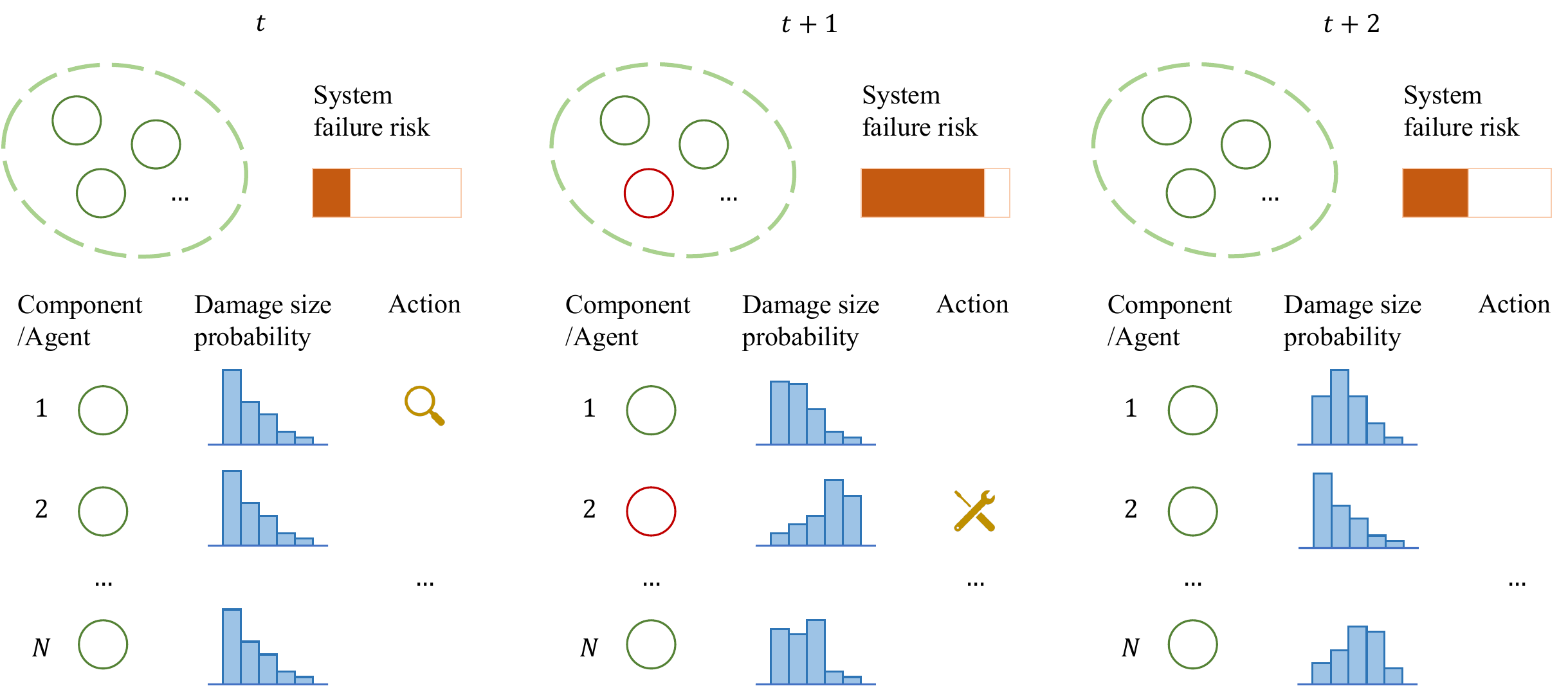}
    \caption{IMP environment~\citep{benchmark_IMP}. This environment simulates an engineering system with multiple components controlled by agents. The objective is to minimize overall system failure risk at low costs. The system risk depends on component damage probabilities, which evolve over time and can be influenced by agent inspection or repair actions.}
    \label{fig: IMP_env}
\end{figure}

\paragraph{MAgent} MAgent is a highly scalable gridworld gaming benchmark shown in \cref{fig: Magent_env}. In \texttt{AdversarialPursuit}, agents aim to tag adversaries while adversaries try to escape. Agents can choose actions from \texttt{move}, \texttt{tag} or \texttt{do nothing}. Agents are rewarded for successfully tagging an adversary and penalized for unsuccessful tagging attempts. In \texttt{Battle}, agents attempt to attack and eliminate adversaries, with the same goal for adversaries. Agents can choose actions from \texttt{move}, \texttt{attack} or \texttt{do nothing}. A team wins by eliminating all opponents or having more surviving agents when the episode ends.

Following the official implementation~\citep{magent2}, we use an individual reward setting with IDQN as the base algorithm. 
Due to the high-dimensional observations in MAgent, storing experience in a replay buffer can be challenging because of hardware constraints. We adopt a preprocessing procedure following \citet{DGN}, compressing observations by concatenating [my\_team\_hp - obstacle/off the map, other\_team\_hp - obstacle/off the map]. To facilitate the use of communication, we use small view ranges ($8$ for \texttt{AdversarialPursuit} and $7$ for \texttt{Battle}). In both scenarios, we pretrain the adversary policies using the IDQN algorithm with a self-play scheme and use these pretrained policies to test the performance of different algorithms.

\paragraph{IMP} IMP is a platform for benchmarking the scalability of cooperative MARL methods in real-world engineering applications, as illustrated in \cref{fig: IMP_env}. This environment simulates an infrastructure management planning problem with agents controlling different components. Agents can choose actions from \texttt{inspection}, \texttt{repair} or \texttt{do nothing}. In different scenarios, the correlation between agent deterioration processes and the system failure function are defined differently, posing unique challenges.

Following the official implementation of IMP, we use a global reward setting and choose QMIX as the base algorithm due to its stable performance across scenarios. We adopt the campaign cost setting, which requires higher cooperation among agents. As recommended by \citet{benchmark_IMP}, results are normalized with respect to expert-based heuristic policies using $(x-H)/|H|$, where $x$ is the discounted rewards of the tested algorithm, and $H$ is the discounted rewards achieved by heuristic policies listed in \cref{tab: heuristic}.

\begin{table}[th]
    \centering
    \caption{Heuristic policies performance on the IMP benchmark.}
    \label{tab: heuristic}
    \begin{tabular}{ccc}
    \toprule
    Scenario & Number of agents $N$ & Discounted reward $H$ \\
    \midrule
    \multirow{2}{*}{Uncorrelated} & $50$ &  $-232.7$ \\
    & $100$ & $-231.5$ \\
    \midrule
    \multirow{2}{*}{Correlated} & $50$ &  $-211.0$ \\
    & $100$ & $-194.0$ \\
    \midrule
    \multirow{2}{*}{OWF} & $50$ &  $-1248.2$ \\
    & $100$ & $-2436.3$ \\
    \bottomrule
    \end{tabular}

\end{table}

\subsection{Implementation details} \label{supp: implement_details}
In MAgent, we implement our proposed ExpoComm along with baselines DGN+TarMAC and ER on top of the IDQN base algorithm. In IMP, these are implemented on top of QMIX. For ExpoComm, we use \cref{eq: aux_pred} for MAgent benchmark because the global state is provided in this environment,  and \cref{eq: aux_cont} for IMP, as the global state is a concatenation of all observations and is not compact or suitable for message grounding.

\subsection{Experimental Infrastructure} \label{supp: infrast}
The experiments were conducted using NVIDIA GeForce RTX 3080 GPUs and NVIDIA A100GPUs. Each experimental run required less than 2 days to complete.

\section{More results and discussion}

\subsection{Visualization results} \label{supp: vis_results}

\begin{figure}[t]
\centering
\begin{subfigure}[t]{0.8\textwidth}
    \centering
    \includegraphics[width=\textwidth]{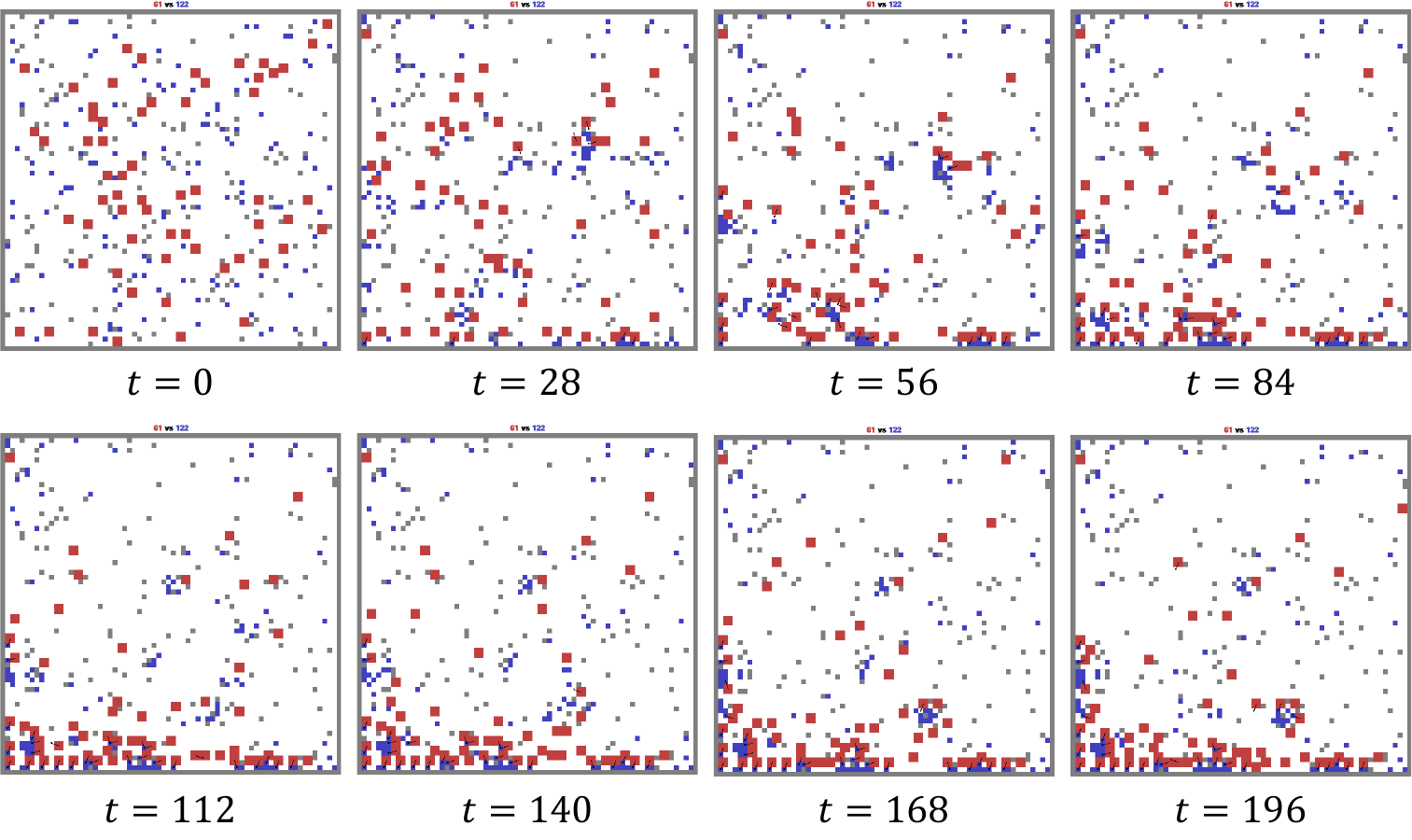}
    \caption{ExpoComm policies with $K=1$. Red agents push blue adversary agents to the edge of the frame and trap them there to obtain high rewards by repeatedly tagging them.}
    \label{fig: vis_adv_pursuit_ExpoComm}
\end{subfigure}
\begin{subfigure}[t]{0.8\textwidth}
    \centering
    \includegraphics[width=\textwidth]{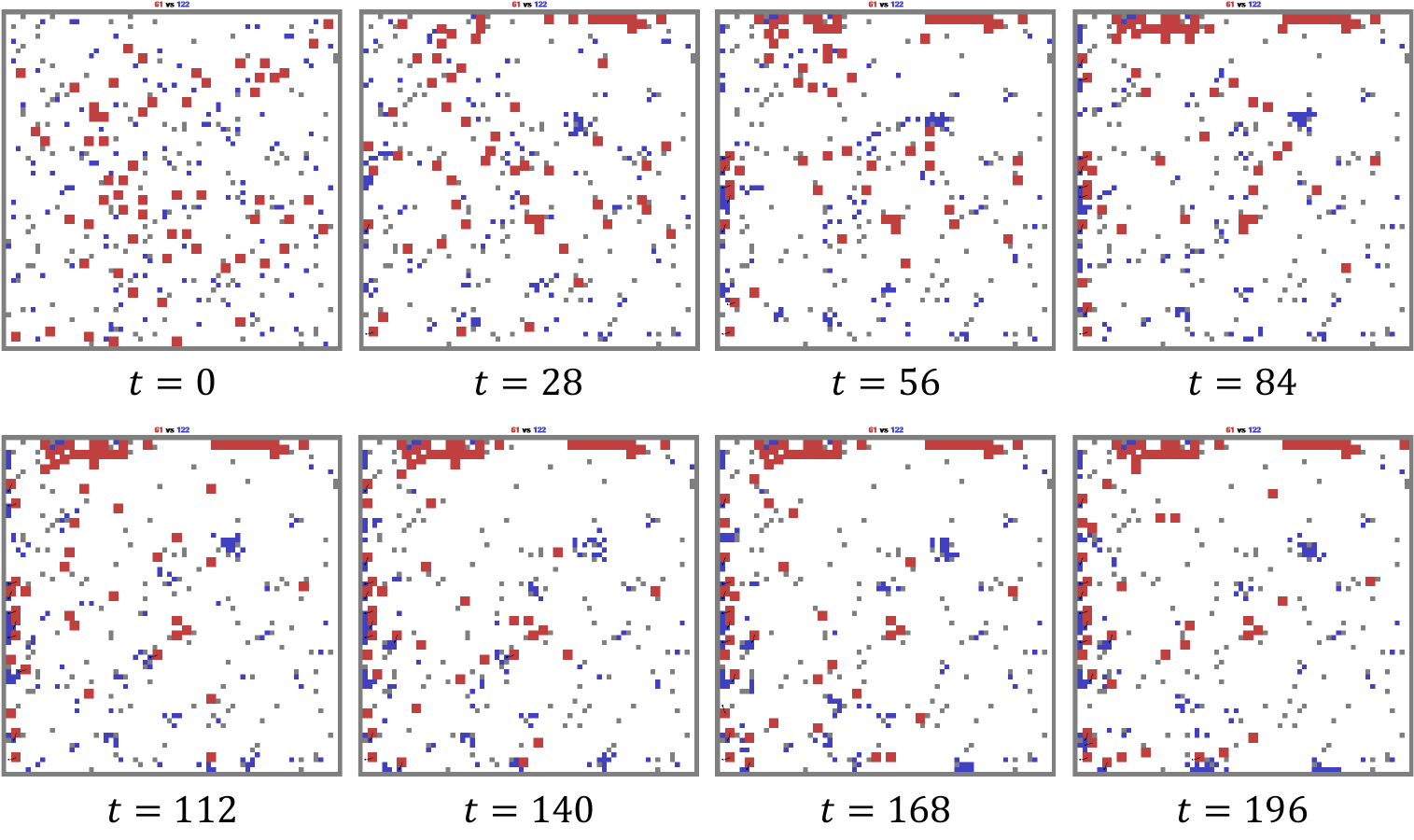}
    \caption{IDQN policies.}
    \label{fig: vis_adv_pursuit_IDQN}
\end{subfigure}
\caption{Visualization in AdversarialPursuit w/ 61 agents.}
\label{fig: vis_adv_pursuit}
\end{figure}

\begin{figure}[t]
\centering
\begin{subfigure}[t]{0.8\textwidth}
    \centering
    \includegraphics[width=\textwidth]{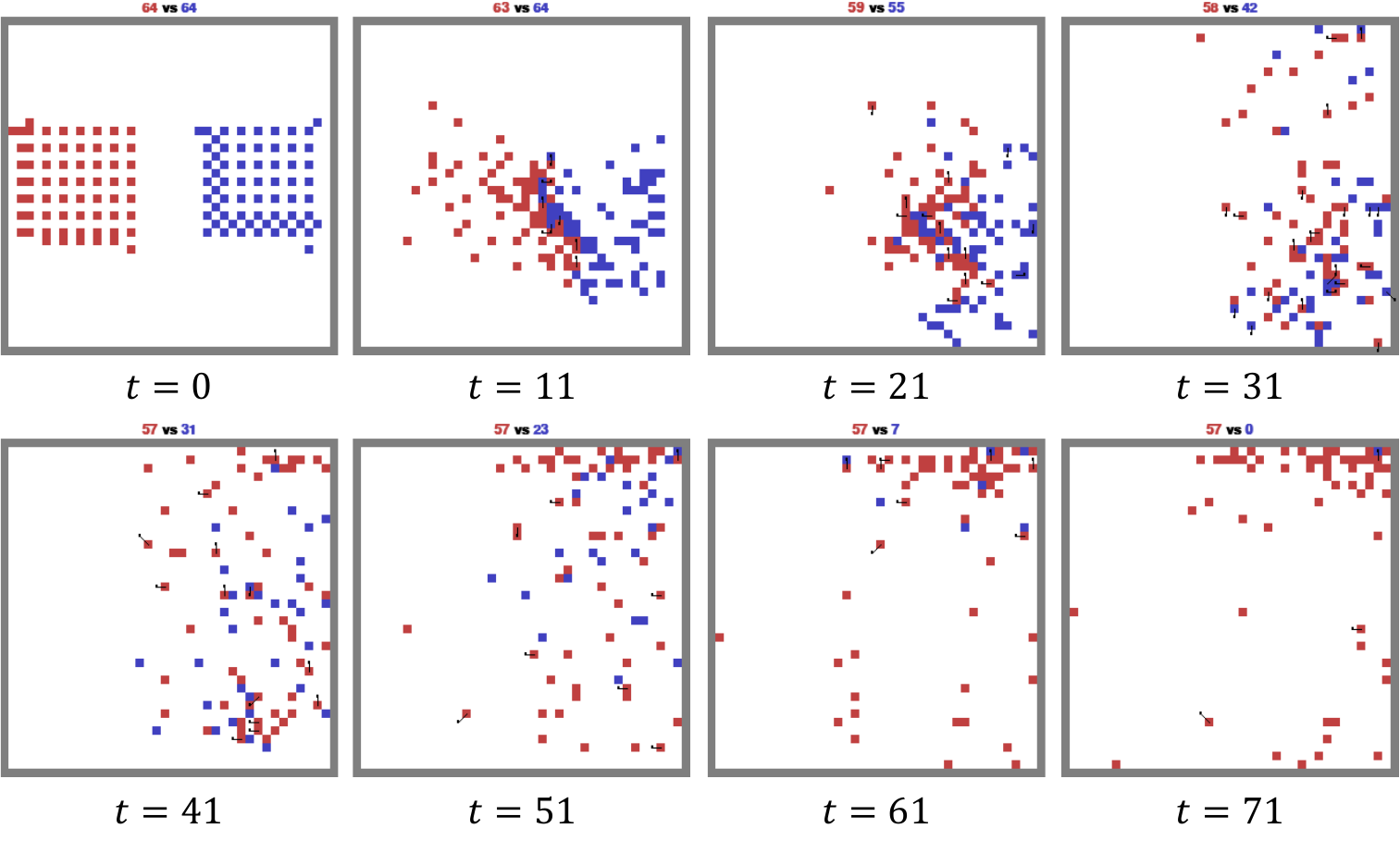}
    \caption{ExpoComm policies with $K=1$. Agents coordinate to ensure red agents outnumber blue adversaries on the front line ($t=11$, $t=21$), securing an advantage. Once red agents substantially outnumber blue adversaries, they surround the remaining adversaries ($t=61$, $t=71$) to eliminate them.}
    \label{fig: vis_battle_ExpoComm}
\end{subfigure}
\begin{subfigure}[t]{0.8\textwidth}
    \centering
    \includegraphics[width=\textwidth]{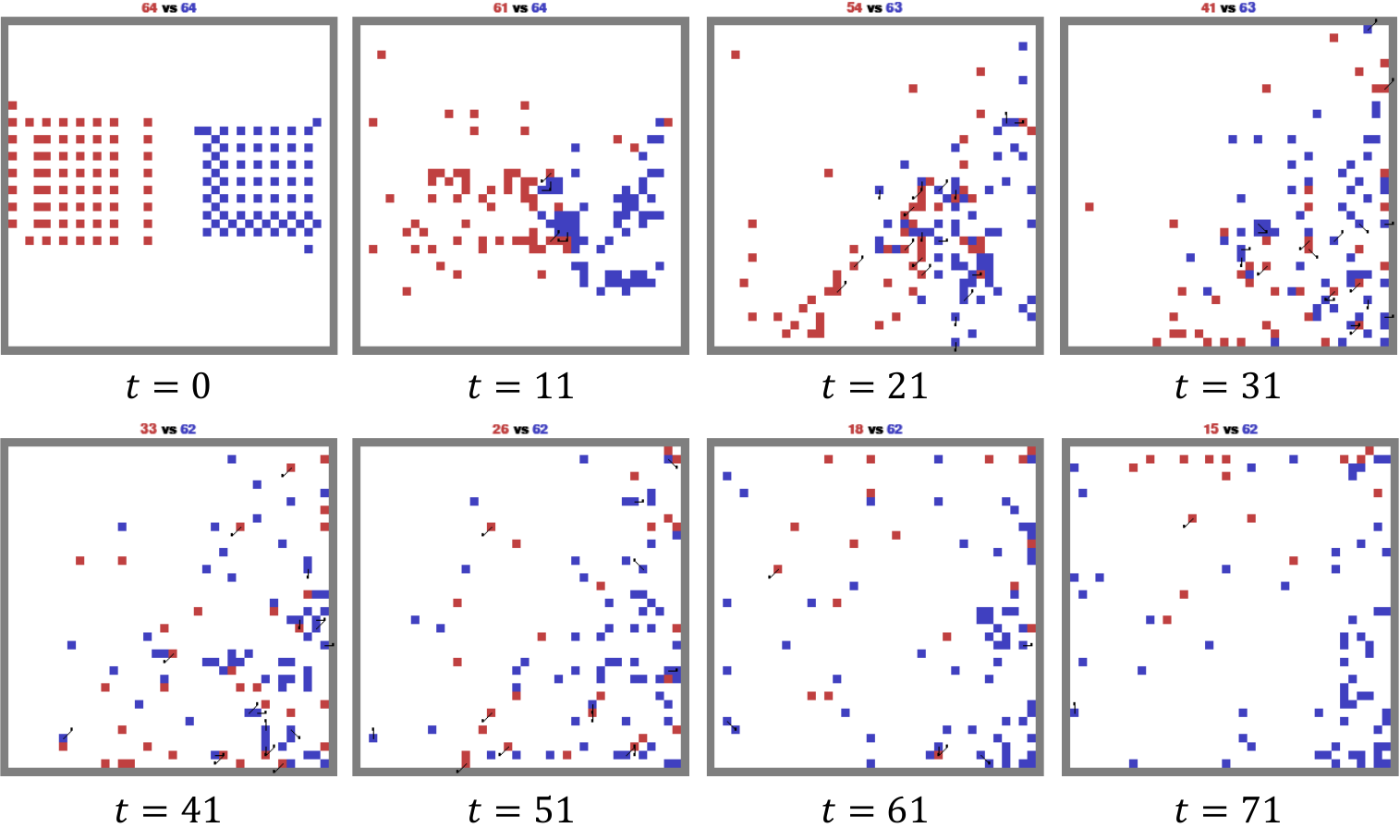}
    \caption{IDQN policies.}
    \label{fig: vis_battle_IDQN}
\end{subfigure}
\caption{Visualization in Battle w/ 64 agents.}
\label{fig: vis_battle}
\end{figure}

We visualize the final trained policies of ExpoComm and IDQN in \texttt{dversarialPursuit} and \texttt{Battle} with \cref{fig: vis_adv_pursuit} and \cref{fig: vis_battle} respectively to demonstrate how ExpoComm enhances cooperation among agents. As shown in \cref{fig: vis_adv_pursuit_ExpoComm} and \cref{fig: vis_battle_ExpoComm}, agents adopt a global perspective and act cooperatively with ExpoComm policies, demonstrating effectiveness even under extreme communication budgets($K=1$). In comparison, IDQN agents focus only on local observations and often become trapped in suboptimal solutions due to lack of coordination.

\subsection{Comparison with proxy-based communication}

\begin{figure}[t]
\centering
% \subfigure[]{
% \includegraphics[width=0.7\textwidth]{Styles/figs/results/0515_legend_cropped.pdf}
%     }8
\raisebox{-\height}{\includegraphics[width=0.75\textwidth]{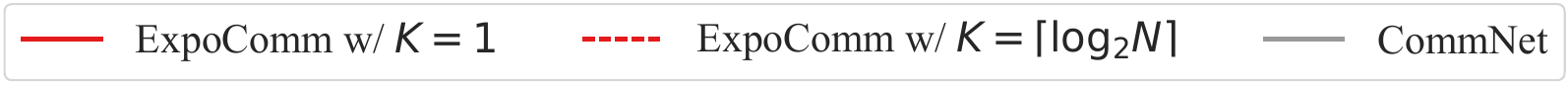}}
\par
\begin{subfigure}[t]{.32\textwidth}
    \centering
    \includegraphics[width=\textwidth]{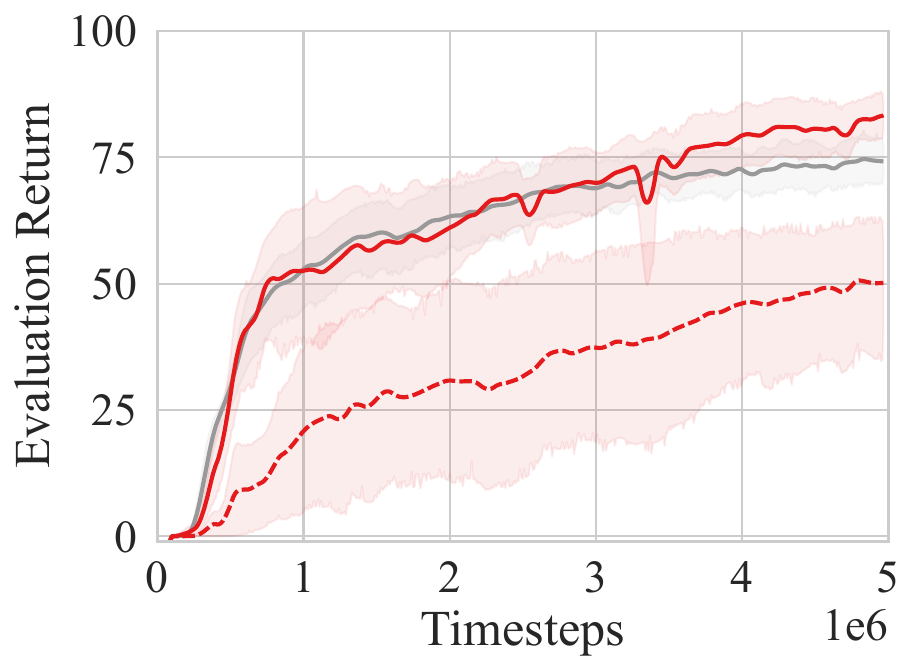}
    \caption{AdversarialPursuit w/ 25 agents}
\end{subfigure}
\begin{subfigure}[t]{.32\textwidth}
    \centering
    \includegraphics[width=\textwidth]{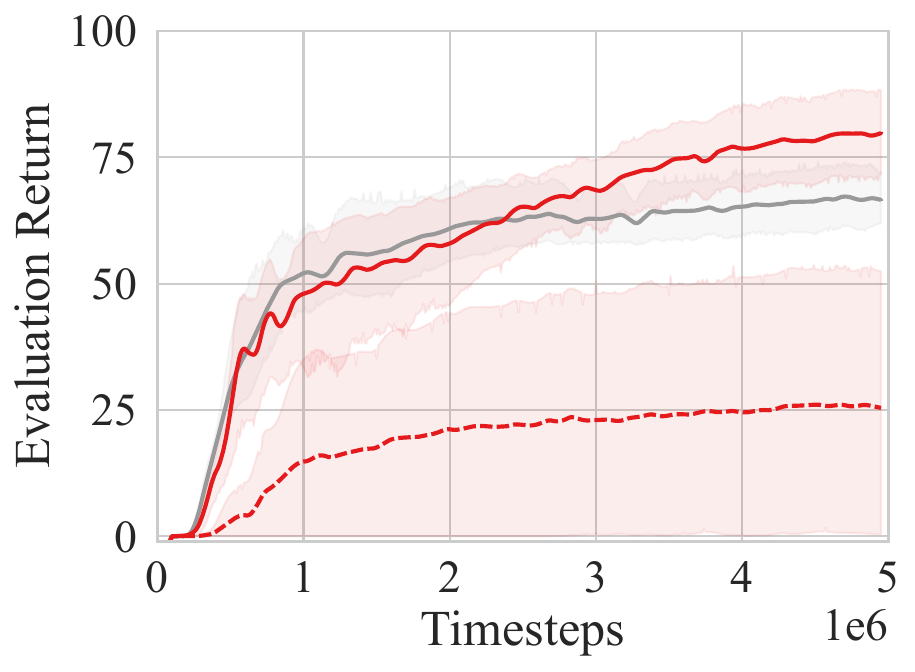}
    \caption{AdversarialPursuit w/ 45 agents}
\end{subfigure}
\begin{subfigure}[t]{.32\textwidth}
    \centering
    \includegraphics[width=\textwidth]{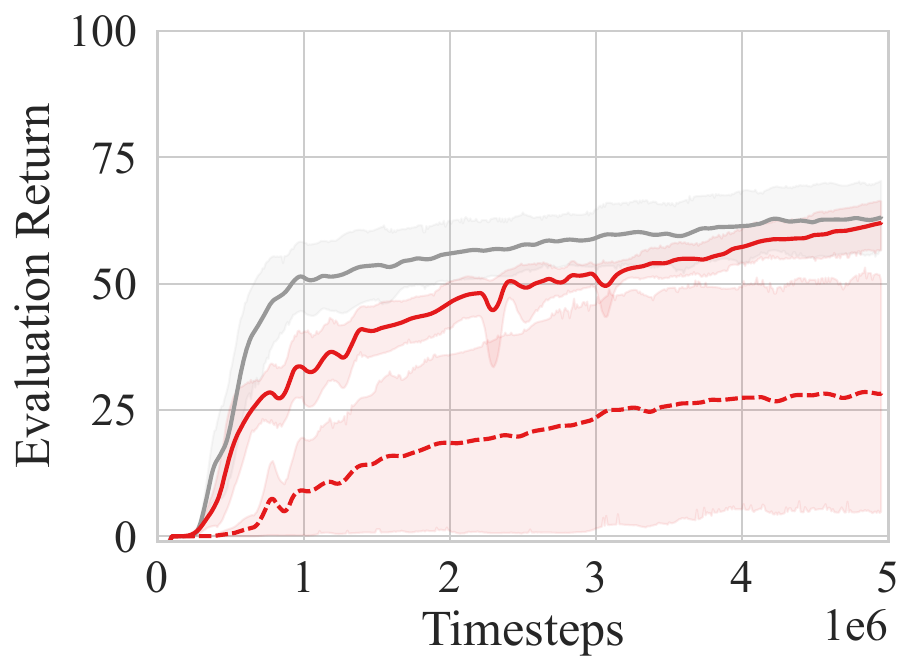}
    \caption{AdversarialPursuit w/ 61 agents}
\end{subfigure}
\begin{subfigure}[t]{.32\textwidth}
    \centering
    \includegraphics[width=\textwidth]{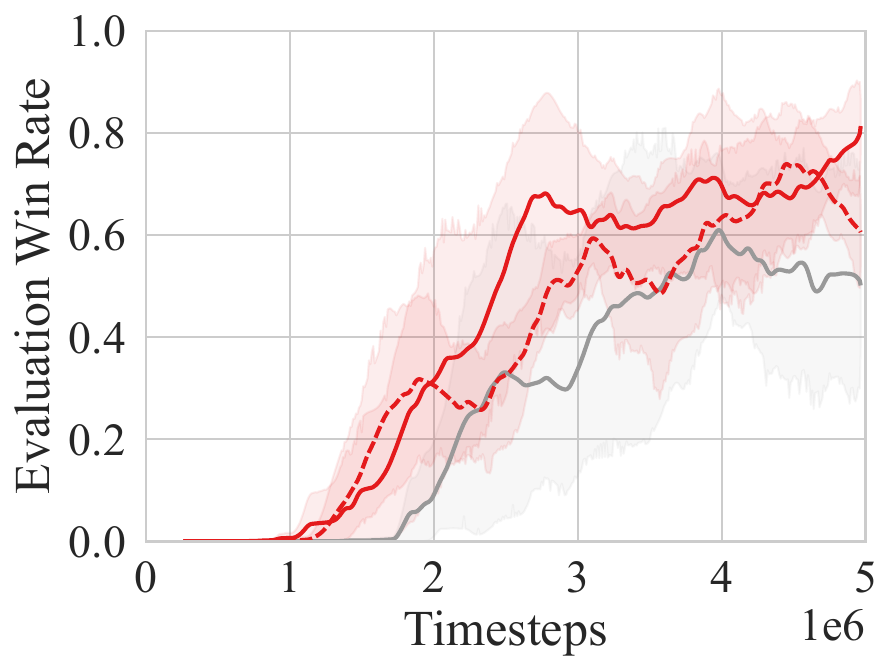}
    \caption{Battle w/ 20 agents}
\end{subfigure}
\begin{subfigure}[t]{.32\textwidth}
    \centering
    \includegraphics[width=\textwidth]{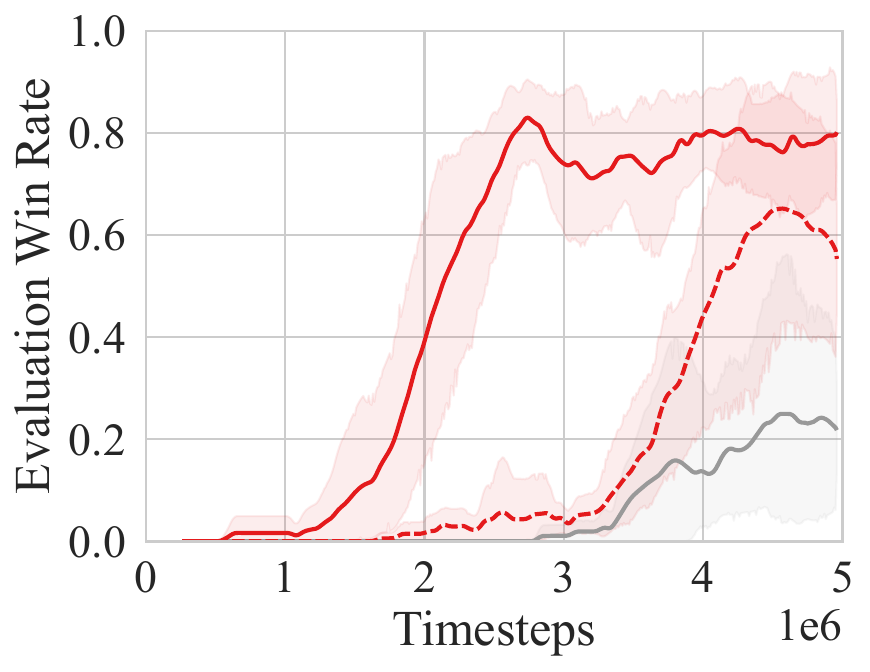}
    \caption{Battle w/ 42 agents}
\end{subfigure}
\begin{subfigure}[t]{.32\textwidth}
    \centering
    \includegraphics[width=\textwidth]{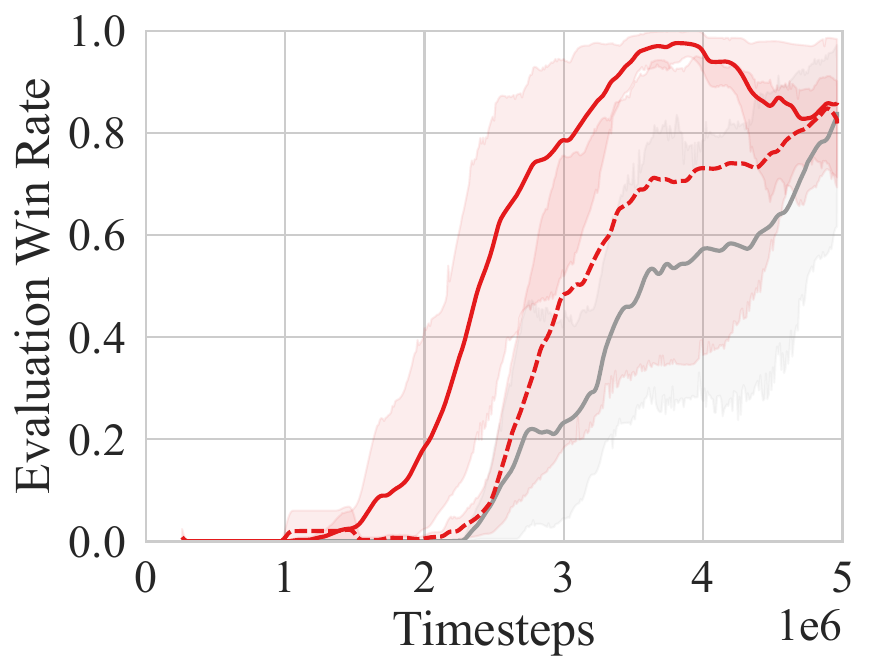}
    \caption{Battle w/ 64 agents}
\end{subfigure}
\caption{Performance comparison with proxy-based baselines on MAgent tasks.}
\label{fig:results_magent_CommNet}
\end{figure}

Although we primarily focus on decentralized communication-based MASs without centralized proxies, we also compare ExpoComm against the proxy-based CommNet~\citep{CommNet}. As seen in \cref{fig:results_magent_CommNet} and \cref{tab:results_imp_CommNet}, ExpoComm outperforms CommNet in most scenarios, especially in IMP benchmarks. However, CommNet achieves comparable performance on the \texttt{AdversarialPursuit} tasks. This implies that a global perspective is more crucial for success in these scenarios, possibly explaining ExpoComm's larger advantage over other baselines in this scenario.

\begin{table}[t]
\centering
\begin{threeparttable}
\caption{Performance comparison with proxy-based baselines on IMP tasks. Results are reported as the mean and standard deviation of the percentage of normalized discounted rewards relative to expert-based heuristic policies, following \citet{benchmark_IMP}, with details in \cref{supp: env_details}. The best-performing method is indicated in \textbf{bold}, and the second best is \underline{underlined}.}
\label{tab:results_imp_CommNet}
\begin{tabular}{lccc}
\toprule
\multirow{2}{*}{Scenario} & \multicolumn{1}{c}{CommNet} & \multicolumn{2}{c}{ExpoComm} \\
\cmidrule(lr){2-2} \cmidrule(lr){3-4}
& with communication proxy & $K=1$ & $K=\lceil\log_2N\rceil$ \\
\midrule
\multicolumn{4}{c}{$N=50$} \\
\midrule
Uncorrelated & $26.07\,\scalebox{0.8}{($6.82$)}$ & $\underline{27.31\,\scalebox{0.8}{($2.26$)}}$ & $\mathbf{28.26\,\scalebox{0.8}{($2.51$)}}$ \\
Correlated & $26.14\,\scalebox{0.8}{($16.87$)}$ & $\mathbf{43.82\,\scalebox{0.8}{($6.33$)}}$ & $\underline{40.01\,\scalebox{0.8}{($3.19$)}}$ \\
OWF & $53.71\,\scalebox{0.8}{($1.27$)}$ & $\underline{64.66\,\scalebox{0.8}{($0.26$)}}$ & $\mathbf{65.19\,\scalebox{0.8}{($0.51$)}}$ \\
\midrule
\multicolumn{4}{c}{$N=100$} \\
\midrule
Uncorrelated & $-65.92\,\scalebox{0.8}{($125.03$)}$ & $\underline{27.34\,\scalebox{0.8}{($13.32$)}}$ & $\mathbf{27.81\,\scalebox{0.8}{($5.71$)}}$ \\
Correlated & $-82.76\,\scalebox{0.8}{($48.62$)}$ & $\mathbf{19.17\,\scalebox{0.8}{($23.94$)}}$ & $\underline{17.25\,\scalebox{0.8}{($22.70$)}}$ \\
OWF & $34.71\,\scalebox{0.8}{($5.34$)}$ & $\underline{65.26\,\scalebox{0.8}{($1.34$)}}$ & $\mathbf{66.23\,\scalebox{0.8}{($0.38$)}}$ \\
\bottomrule
\end{tabular}
\end{threeparttable}
\end{table}

\subsection{Limitations and future work}
While ExpoComm demonstrates strong performance and scalability in cooperative multi-agent tasks, some limitations remain.

First, ExpoComm does not explicitly incorporate agent heterogeneity or properties of the underlying environmental MDP when constructing the communication topology. This could result in suboptimal performance in scenarios requiring targeted messaging between specific agents~\citep{MAIC} or in networked MDPs~\citep{networked_MA, large_scale_networked_MARL}. Therefore, Incorporating factors like agent identities or relationships presents a promising direction for further improvements in such settings.

Second, we evaluated ExpoComm primarily in fully cooperative tasks. Partially competitive settings requiring agents to learn to communicate only when necessary remain challenging. Examining ExpoComm's capabilities and limitations in such partially competitive tasks presents an important avenue for future work.

Finally, communication scalability in multi-agent systems remains an under-explored area despite the attempt of this work. For instance, incorporating finer graph topologies beyond exponential graphs may enhance performance, and exploiting temporal communication sparsity could further reduce costs. There are still many open questions in scaling communication efficiently.

\end{document}